\documentclass[%
reprint,
amsmath,amssymb,
aps,
pra,
]{revtex4-2}

\usepackage{graphicx}
\usepackage{dcolumn}
\usepackage{bm}
\usepackage{empheq} 
\usepackage{fancybox} 
\usepackage{longtable}

\usepackage{tcolorbox}
\tcbuselibrary{theorems}

\usepackage[english]{babel}
\makeatletter
\let\ORIbbl@fixname\bbl@fixname
\def\bbl@fixname#1{%
	\@ifundefined{languagealias@\expandafter\string#1}
	{\ORIbbl@fixname#1}
	{\edef\languagename{\@nameuse{languagealias@#1}}}%
}
\newcommand{\definelanguagealias}[2]{%
	\@namedef{languagealias@#1}{#2}%
}
\makeatother
\definelanguagealias{en}{english}
\definelanguagealias{EN}{english}
\definelanguagealias{eng}{english}
\definelanguagealias{fr}{english}

\begin{document}

\title{Combs, fast and slow: non-adiabatic mean field theory of active cavities}

\author{David Burghoff}
\affiliation{%
 Chandra Family Department of Electrical and Computer Engineering, University of Texas at Austin, Austin, TX 78712
}%

%
%

\date{\today}

\begin{abstract}
Integrated frequency combs based on active cavities are of interest for a wide range of applications. An elegant description of these cavities is based on mean-field theory, which averages the effect of internal dynamics occurring within a round trip. Lasers based on media with slow gain dynamics can be described by solving the population over many round trips, while lasers based on fast gain media can be described by adiabatic elimination. However, most gain media actually have both fast and slow components, and effects often ascribed to fast gain media are known to arise even in slower gain media. Here, we develop an operator-based mean-field theory that non-adiabatically describes the dynamics of bidirectional active cavities, both fast and slow. It is based on first principles and semi-exactly replaces the Maxwell-Bloch equations, but is flexible enough to accomodate non-trivial lineshapes and population dynamics. As an example, we use this formalism to establish an additional constraint on the formation of frequency-modulated combs. Our results are general and apply to any bidirectional or unidirectional active cavity, and as a result, generalize to essentially any chip-scale laser.

\end{abstract}

\pacs{Valid PACS appear here}
\maketitle


\newcommand{\nlgainvar}{\Delta} 

\section{\label{sec:level1}Introduction}

Lasers are some of the richest dynamical systems to have ever been produced and still produce surprises. Over the last few years, a number of comb states have been shown to appear in systems with fast gain media whose population dynamics occur much faster than their round-trip time. Perhaps chief among these is the linearly-chirped frequency-modulated (FM) comb state, which was first conclusively identified in quantum cascade lasers using SWIFTS \cite{singletonEvidenceLinearChirp2018,burghoffEvaluatingCoherenceTimedomain2015}, but has also been shown to appear in a wide range of semiconductor systems \cite{sterczewskiFrequencymodulatedDiodeLaser2020,daySimpleSinglesectionDiode2020,hillbrandInPhaseAntiPhaseSynchronization2020, dongBroadbandQuantumdotFrequencymodulated2023,krisoSignaturesFrequencymodulatedComb2021} using a variety of experimental techniques \cite{taschlerFemtosecondPulsesMidinfrared2021,cappelliRetrievalPhaseRelation2019,heckPassivelyModeLocked102009}. Other comb states relying on the fast nature of the gain medium include the quantum walk state \cite{heckelmannQuantumWalkComb2023}, the Nozaki-Bekki hole \cite{opacakNozakiBekkiSolitons2024}, and the production of bright-dark pairs in coupled rings \cite{letsouLasingHybridizedSoliton2024}. When the gain medium's population dynamics are fast relative to the round trip time, the population and coherence dynamics can be adiabatically eliminated \cite{opacakTheoryFrequencyModulatedCombs2019}, permitting analytical solutions to be found \cite{burghoffUnravelingOriginFrequency2020}. This is in stark contrast to the usual description of slow gain media, which are typically described by the Haus master equations that eliminate the dynamics entirely \cite{hausTheoryModeLocking1975}. But how fast is fast?

The most complete description of lasers above threshold are the Maxwell-Bloch equations. These are completely exact and can be solved numerically, but usually cannot be solved analytically except in the case of very specific trivial solutions. Moreover, the parameter space is large, and the specific gain dynamics of the laser must be included---meaning one is subjected to an enormous number of parameters that in many cases are not even measurable. It is for this reason that novel states are mostly driven by experiment. For example, even though the linearly-chirped FM state is now recognized as a somewhat universal state in semiconductor lasers, it had not been predicted at all.

A more recent development has been the creation of mean-field theories that are analogous to the famed Lugiato-Lefever equation \cite{lugiatoSpatialDissipativeStructures1987}. When the internal dynamics of a field within a round-trip are small, these changes can be averaged and ignored. The Lugiato-Lefever equation has been known for several years to describe Kerr combs effectively \cite{coenModelingOctavespanningKerr2013,chemboSpatiotemporalLugiatoLefeverFormalism2013}, but was not applied semiconductor lasers since they typically have large outcoupling losses (violating the main assumption of mean-field theory). However, we showed in \cite{burghoffUnravelingOriginFrequency2020} that by normalizing the field to its continuous-wave (CW) profile, one could construct a mean-field theory that applies to lasers---essentially the LLE for active cavities. When applied to a bidirectional Fabry-Perot cavity it can result in dynamics governed by a phase-driven nonlinear Schrodinger equation, and when applied to a unidirectional cavity it becomes a complex Ginsburg-Landau equation \cite{columboUnifyingFrequencyCombs2021}.

The chief technique that has been used to describe fast gain media analytically has been the adiabatic approximation \cite{opacakTheoryFrequencyModulatedCombs2019,burghoffUnravelingOriginFrequency2020,senicaFrequencyModulatedCombsFieldEnhancing2023,opacakNozakiBekkiSolitons2024,heckelmannQuantumWalkComb2023}. One can show from the Bloch equations that the population and coherence are each related to the quantity $(1+T_i \partial_t)^{-1}$, where $T_1$ is the population inversion lifetime and $T_2$ the coherence lifetime. In the adiabatic approximation, the population inversion and coherence are eliminated by taking the low-frequency limit, computing $(1+T_i \partial_t)^{-1} \ \approx 1 - T_i \partial_t+...$, and only taking the first term or two. While adequate in the limit of very fast gain media, it has some troubling aspects. The first is that the geometric series intrinsic to the adiabatic approximation is not convergent for $\omega T_i > 1$. Thus, frequencies too far from the carrier frequency cannot possibly be well-described by it. Secondly, it is numerically poorly-behaved. Such a series generates high frequency components that can cause numerical divergence. As a result, it cannot be applied to systems whose bandwidth is much greater than $1/T_1$ or $1/T_2$. 

Here, we exactly solve the Maxwell-Bloch equations in the non-adiabatic limit, producing both exact normalized master equations and a semi-exact mean-field theory. By introducing Lorentzian operators, we rigorously show that the dynamics of an arbitrary bidirectional cavity can be solved using either
\begin{align}
	\frac{1}{v_g} \frac{\partial f}{\partial t} + \frac{\partial f}{\partial z} &= \frac{1}{2} \hat{g}_{\textrm{NL}} f \quad \text{or}\\
	\frac{1}{v_g} \frac{\partial f}{\partial T} &= \frac{1}{2} \langle \hat{g}_{\textrm{NL}} \rangle f,
\end{align}
where the first equation represents an exact master equation in the normalized field $f$ in terms of the nonlinear gain $\hat{g}_{\textrm{NL}}$ and the group velocity $v_g$, and the latter represents a semi-exact mean-field theory. Either approach that can be solved extremely efficiently using standard split-step methods. Moreover, because our approach is operator-based, different lineshapes and different population dynamics can be included without any additional analysis. With this approach, we demonstrate an additional constraint on FM combs not contained within the adiabatic extendon theory---that there is a maximum limit to the gain recovery time in which extendons cannot form. We also introduce multi-component mean-field theory, which allows for the parsimonious modeling of gain media with both fast and slow components.

\section{Exact master equations and mean-field theory}

The detailed first-principles derivation is given in the appendix, and here we focus on the key results. First, we use the Maxwell-Bloch equations to derive our exact theory. Starting from the evolution of a two-level system with a population lifetime $T_1$ and a coherence lifetime $T_2$, we introduce the operators
\begin{align}
	\hat{L}_i \equiv \left( 1 + T_i \frac{\partial}{\partial t} \right)^{-1}
\end{align}
in order to capture all of the long-term and short-term dynamics. These can be intepreted as Lorentzians in frequency space, since $L_i(\omega)=(1+i\omega T_i)^{-1}$. Using this, we then show in Appendices \ref{exactpol} and \ref{exactmaster} that a field $E_f = A e^{i \omega_0 t} + A^*e^{-i \omega_0 t}$ gives rise to a nonlinear gain of the form
\begin{align} \label{nlgain1}
	\hat{g}_\text{NL} = g_0 \hat{L}_2 \left[ 1 + \frac{1}{P_s} \operatorname{Re} \hat{L}_1 A^* \hat{L}_2 A \right]^{-1} 
\end{align}
where $P_s \equiv E_s^2$ is the saturation field squared and $g_0$ is the small-signal gain. When the usual limits are taken for slow gain, $\hat{L}_2 \rightarrow L_2(\omega)$ and $\hat{L}_1 \rightarrow \hat{I}$, so one recovers the standard form of gain saturation in terms of a homogeneously-broadened lineshape. Still, this expression clarifies that the lineshape operator applies to the product of the field and gain saturation.

The previous expression is valid at a particular point in space, but in most integrated cavities it is possible to have bidirectional fields. In this case, the envelope should be expressed as a sum of counterpropagating waves as $A(z,t) =E_+(z,t) e^{-i k_0 z} + E_-(z,t) e^{i k_0 z}$. Inserting this form into (\ref{nlgain1}) generates an infinite number of spatial Fourier components (at every multiple of $k_0$), but only those terms at $\pm k_0$ will propagate. By calculating the Fourier coefficients and using a standard master equation formalism \cite{opacakTheoryFrequencyModulatedCombs2019}, we find in Appendix \ref{exactmaster} the exact master equations
\begin{align} \label{meqexact}
	\frac{1}{v_g} \frac{\partial E_\pm}{\partial t} &\pm \frac{\partial E_\pm}{\partial z} = \frac{g_0}{2} \hat{L}_2  \frac{1}{\scalebox{1}{$\sqrt{D_0^2 - |D_2|^2}$}} \bigg( E_\pm \nonumber \\
	&+E_\mp \frac{1}{D_2^{\pm *}} \left(\scalebox{1}{$\sqrt{D_0^2 - |D_2|^2} - D_0$} \right) \bigg) \equiv \nlgainvar_\text{NL}^\pm 
\end{align}
where we have introduced the Fourier coefficients 
\begin{align}
	&\quad D_0 \equiv 1 + P_s^{-1} \operatorname{Re} \hat{L}_1 \left( E_+^* \hat{L}_2 E_+ + E_-^* \hat{L}_2 E_- \right) \\
	&\quad D_2^\pm \equiv P_s^{-1} \hat{L}_1 \left( E_\pm (\hat{L}_2 E_\mp)^* + E_\mp^* \hat{L}_2 E_\pm \right)
\end{align}
that are responsible for gain saturation and backscattering, respectively. This form contains all prior adiabatic results \cite{opacakTheoryFrequencyModulatedCombs2019,burghoffUnravelingOriginFrequency2020,peregoCoherentMasterEquation2020} (and indeed, the entire Maxwell-Bloch equations) without any approximation. In the low-field limit, it reduces to
\begin{align}
\nlgainvar_\text{NL}^\pm \approx \frac{g_0}{2} \hat{L}_2 \frac{1}{D_0} \left( E_\pm - \frac{1}{2} E_\mp D_2^{\pm} \right),
\end{align}
demonstrating more explicitly the interpretation of $D_0$ and $D_2$ as respectively creating gain saturation and backscattering from the gain grating. In Fabry-Perot cavities, the backscattering contribution alone is responsible for FM comb generation, as it is the only term that gives rise to phase potentials. In unidirectional cavities, it reduces to 
\begin{align}
	\nlgainvar_\text{NL}^+ &= \frac{g_0}{2} \hat{L}_2  \left(1 + \frac{1}{P_s} \operatorname{Re} \hat{L}_1  E_+^* \hat{L}_2 E_+ \right)^{-1} E_+,
\end{align}
which again resembles familiar intensity saturation. Note also that both Fourier coefficients contain a factor of $L_1$, which can be construed as a low-pass filter in the case of slower gain media.

To construct a mean-field theory, one must account for the primary difference between active cavities and passive cavities: the size of their internal gain and losses. In passive high-Q cavities, the losses and nonlinear gains are typically very small. That is not the case in active cavities---a cleaved semiconductor Fabry-Perot laser, for example, typically has facet reflectivity of just 30\%. This violates the core requirement that the changes be small. In addition, this creates significant explicit position dependence. However, as we previously noted in \cite{burghoffUnravelingOriginFrequency2020}, the large changes can be eliminated by noticing that the power dynamics of a CW laser above threshold are strikingly similar and simple to find (as they are only functions of position). By normalizing to the CW power, the wave becomes periodic over a round trip and mean-field theory can be created.

In Appendix \ref{extnorm}, we perform this normalization. First, we extend the cavity in position space, placing the negative propagating wave at negative z (assuming the cavity coordinates are at positive z). This reduces the master equations to dependence on a single normalized field, $E$. Next, we compute the normalized CW field squared by solving for it in the presence of saturated gain
\begin{align}
	\frac{\partial P}{\partial z} &= \left(-\alpha(z) + g_{cw}(z)\right) P \\
	g_{cw}(z) &\equiv g_0 \frac{P_s}{2P} \left( \scalebox{1.2}{$1+\frac{P - P_- - P_s}{\sqrt{(P + P_- + P_s)^2 - 4 P P_-}}$} \right) 
\end{align}
where $\alpha(z)$ contains all losses (including outcoupling loss), we have introduced $P_-\equiv P(-z)$ as the counterpropagating wave in extended cavity space, and $g_{cw}$ is the saturated gain experienced by CW light. We also construct a dimensionless form of $P$ by dividing it by its saturation value, i.e. $k\equiv P/P_s$. If we then define the normalized extended field as \(f(z,t)=E(z,t)/\sqrt{P_s k(z)}\), we have essentially moved all of the explicit position dependence and large internal dynamics into \(k\) rather than the field. Thus, we arrive at the exact normalized and extended master equation,
\begin{align} \label{exactnormmain}
	&\frac{1}{v_g} \frac{\partial f}{\partial t} + \frac{\partial f}{\partial z} = - \frac{g_{cw}(z)}{2} f + \frac{g_0}{2} \hat{L}_2 \frac{1}{\scalebox{0.9}{$\sqrt{D_0^2 - |D_2|^2}$}} \bigg( f \nonumber \\
	&\quad\quad\quad\quad\quad\quad + \frac{f_-}{2 k s_{-+}^*} \left( \scalebox{0.9}{$\sqrt{D_0^2 - |D_2|^2}  - D_0$} \right) \bigg) \equiv \frac{1}{2} \hat{g}_{\textrm{NL}} f \nonumber \\
	&\quad\quad  D_0 = 1 + k s_{++} + k_- s_{--} \nonumber \\
	&\quad\quad  D_2 = 2 \sqrt{k k_-} \, s_{-+}
\end{align}
where \(s_{ij} \equiv \frac{1}{2} \hat{L}_1 \left( f_i^* \hat{L}_2 f_j + f_j ( \hat{L}_2 f_i )^*\right)\) are intensity-like. Again, this result is completely exact and encapsulates all dynamics of a two-level Maxwell-Bloch formalism. It is also more amenable to spectral methods than the standard master equations, as it is periodic and translation invariant. For unidirectional waves, the nonlinear gain reduces to
\begin{align}
	&\hat{g}_{\textrm{NL}}= - g_{cw}(z) + \hat{L}_2 \frac{g_0}{1 + k \operatorname{Re} \hat{L}_1 ( f^* \hat{L}_2 f )  }  \nonumber \\
	&\quad g_{cw}(z) = g_0 \frac{1}{1+k}
\end{align}
which again is reminiscent of standard intensity saturation.

We can now use this to construct a mean-field theory. For a nonlinear gain that depends on waves traveling in each direction \(\frac{1}{2} \hat{g}_{\textrm{NL}} f = \nlgainvar(f, f_-)\), we show in Appendix \ref{mfaverage} that an arbitrary mean field theory can be found by computing
\begin{align} 
	\langle \nlgainvar \rangle = \frac{1}{L_r} \int_0^{L_r} \big[ \nlgainvar (  f(z - v_g t), f(-z - v_g t)) \big]_{z \to -u \atop t \to -\frac{u + z}{v_g}} \, du,
\end{align} 
since the normalized field primarily travels in one dimension. While this form is general, when applied to the exact nonlinear gain a difficulty in efficient calculation is that the waves traveling in opposite directions are no longer separable. Thus, we make a low-field approximation that expands (\ref{exactnormmain}) to first order in k (equivalent to a third-order expansion in E):
\begin{align} 
	\Delta = &\frac{g_0}{2} \left( \hat{L}_2 - 1 + k + 2k_- \right) f \nonumber \\
	-&\frac{g_0}{2} \hat{L}_2 \left[ (k s_{++} + k_- s_{--}) f + k_- s_{-+} f_- \right] 
\end{align} 
Note, however, that when the exact version of \(k\) is used, this expansion is actually better-than-third order in the field (correctly predicting the CW power exactly). Thus, we refer to it as semi-exact. Finally, we introduce the position-space version of \(\hat{L}_i\) (denoted by \(\hat{\mathcal{L}}_i\)), the bandpass operators  \(\hat{\mathcal{B}}_{ik}\), and the filtered intensity \(s\) as
\begin{align}
	\hat{\mathcal{L}}_i &\equiv \hat{L}_i(\partial_t \rightarrow -v_g \partial_z) = ( 1 - v_g T_i \partial_z )^{-1} \\
	\hat{\mathcal{B}}_{ik} &\equiv (\hat{\mathcal{L}}^{-1}_i - 1)^k \hat{\mathcal{L}}^{k+1}_i \\
	s &\equiv \operatorname{Re} \hat{\mathcal{L}}_1 \left( f^* \hat{\mathcal{L}}_2 f \right)
\end{align}
allowing us to at last write down a semi-exact mean-field expression for the gain as
\begin{align}
	\frac{1}{v_g} \frac{\partial f}{\partial T} &= \frac{1}{2} \langle \hat{g}_{\textrm{NL}} \rangle f \nonumber\\
	\frac{\langle \hat{g}_{\textrm{NL}} \rangle}{g_0} &= \hat{\mathcal{L}}_2 - \hat{I}  \nonumber - \bigg( \langle k \rangle \hat{\mathcal{L}}_2 s + \sum_{m=0}^{\infty} \tilde{k} [ \hat{\mathcal{B}}_{2m} s ] \hat{\mathcal{B}}_{2m} - 3 \langle k \rangle \bigg) \nonumber \\[-1.5mm]
	&\hspace{1mm} - \frac{1}{2} \sum_{n,m=0}^{\infty} \Big( \tilde{k} [ \hat{\mathcal{B}}_{2m} f \, \hat{\mathcal{B}}_{1n} f^* ]  \hat{\mathcal{L}}_2 \nonumber  \\[-3mm]
	&\hspace{1.55cm}+\tilde{k} [ \hat{\mathcal{B}}_{2m} f \, \hat{\mathcal{B}}_{1n} \hat{\mathcal{L}}_2 f^* ] \Big) \hat{\mathcal{B}}_{2m} \hat{\mathcal{B}}_{1n} 
\end{align}
where we introduce the notation \(\langle k \rangle \equiv \tfrac{1}{L_r} \int_0^{L_r} k(u) du\) and \(\tilde{k} \left[ f \right] (z) \equiv \frac{1}{L_r} \int_0^{L_r} k(u) f(z + 2u) du \), which occur when dealing with self-interaction and counterpropagation interaction, respectively. The first line represents the contribution from the lineshape and from gain saturation, while the second and third represent the backscattering. While we have written it down in terms of the spatial domain, it could instead be written in terms of fast time by using the temporal version of the operators.

\begin{figure}
	\includegraphics[width=\linewidth]{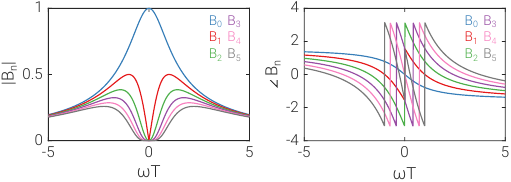}
	\caption{\label{fig:product_series}Product series operators in the frequency domain. Each \(\hat{\mathcal{B}}_{in}\) can be treated as a bandpass filter peaking at \(|\omega T_i| = \sqrt{n}\). At high frequencies, each operator decays to zero, which ensures that the series is well-conditioned.}
\end{figure}

In the above expression, the sums were introduced using an identity proven in Appendix \ref{lorfactor} that distributes \(\hat{L}_i\) over a product and permits the separation of counterpropagating terms. Critically, we also prove that this sum is convergent for \textit{any} choice of frequency, unlike the adiabatic expansion that is intrinsically limited to low frequencies. While the semi-exact model is lengthy, in many regards it is simpler than its adiabatic approximation previously reported (\cite{burghoffUnravelingOriginFrequency2020}), as it replaces the five backscattering terms with just two. Moreover, it has eliminated the various ``magic'' constants that arise from multiple adiabatic expansions (like 3/2 and 5/2), and only the lowest-order \(\hat{\mathcal{B}}_{i0}=\hat{\mathcal{L}}_i\) is needed to replicate the adiabatic theory. This theory clarifies the role and proper ordering of the various time constants, and most importantly decouples the gain and coherence dynamics from the field dynamics. As the operator $g_\text{NL}$ is expressed in terms of $\hat{L}_1$ and $\hat{L}_2$, one can change the lineshape merely by changing $\hat{L}_2$, and one can change the gain dynamics merely by changing $\hat{L}_1$.

\section{Numerical results}

In contrast to the adiabatic theory, which suffers from numerical issues due to the presence of derivatives (which naturally amplify high frequencies), this theory is numerically well-behaved and lacks any explicit derivatives. Higher-order derviatives are instead replaced with higher-order \(\hat{\mathcal{B}}_{in}\) operators. Each operator can be construed as a filter: each \(\hat{\mathcal{B}}_{in}\) represents a bandpass filter peaking at \(|\omega T_i| = \sqrt{n}\) (Figure \ref{fig:product_series}). In the adiabatic theory, phase potentials are generated through the slope of \( \operatorname{Im} \mathcal{B}_{i0}(\omega)\) at its zero-crossing; in the generalized theory, each \(  \mathcal{B}_{in}(\omega)\) can generate phase potentials around a different frequency.

\begin{figure}
	\includegraphics[width=\linewidth]{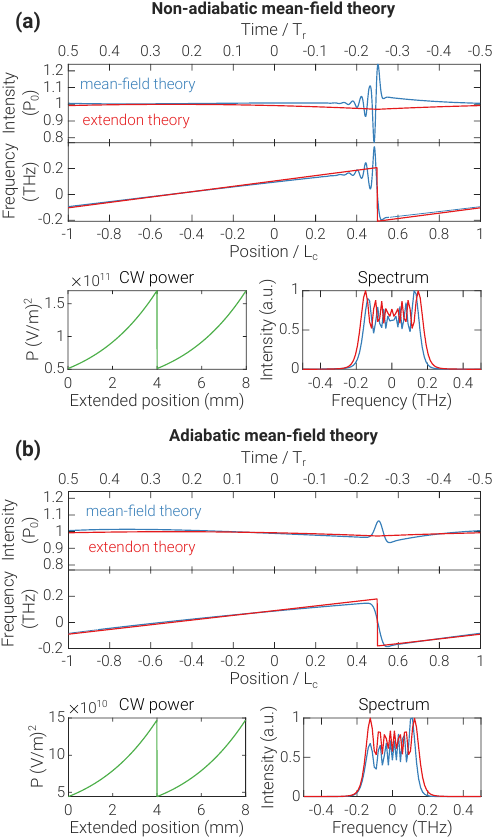}
	\caption{\label{fig:adiabatic_comp}Comparison between the non-adiabatic and adiabatic mean-field theories in the fast gain limit ($T_1=0.2$ ps) with a moderate dispersion  ($k''=$-2000 fs$^2$/mm) (a) Semi-exact non-adiabatic mean-field theory results showing the instantaneous intensity (top), frequency (middle), CW field squared (bottom left), and spectrum (bottom right).  The agreement between the simulation and extendon theory is excellent except at the frequency jump point, correctly predicting the chirp and accurately predicting the spectrum. The theory breaks down at the turnaround point, where an amplitude pulsation develops with noticeable ringing. b. Same dynamics in the low-field, adiabatic limit. The amplitude pulsation at the jump point has much lower ringing, although the predicted spectral shape has an asymmetry.}
\end{figure}

To verify that our analysis agrees with the adiabatic theory in the fast gain recovery limit, we implement it using a standard symmetric split-step Fourier method; these results are shown in Figure \ref{fig:adiabatic_comp}. Figure \ref{fig:adiabatic_comp}a shows the results of the new non-adiabatic mean-field theory, while Figure \ref{fig:adiabatic_comp}b shows the same results using the adiabatic mean-field theory. Both also show the analytical results from the extendon theory \cite{humbardAnalyticalTheoryFrequencymodulated2022}. In both cases, the agreement with the extendon theory is excellent except at the point where the instantaneous frequency jumps. Still, there is some notable difference in the dynamics at the jump point: more ringing is observed in the non-adiabatic case. These are actually a result of the gain recovery dynamics, and are somewhat analagous to relaxation oscillations. Somewhat counterintuitively, the extendon theory is actually in better agreement with the non-adiabatic theory than with the adiabatic theory in terms of the spectral shape, as the non-adiabatic theory predicts an even flatter intensity away from the jump.

\begin{figure}
	\includegraphics[width=\linewidth]{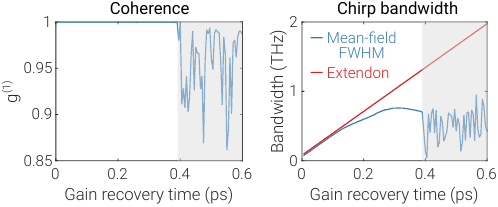}
	\caption{\label{fig:gr_sweep}Variation of the non-adiabatic theory as the gain recovery time is swept and the small-signal gain is held constant. Coherence is shown on the left and chirp bandwidth on the right for a comb with a dispersion of $k''=$-1000 fs$^2$/mm. The unstable region is indicated by the shaded region.}
\end{figure}

Next, we show the effect of changing gain recovery time in Figure \ref{fig:gr_sweep}. Since the small-signal gain is proportional to $T_1$ and is more properly thought of as a requirement of the cavity, it is held constant to a value above threshold. Similarly, $P_s$'s dependence on $T_1$ is not relevant to any dynamics since the normalized theory does not depend on it at all---it contributes only the final scale factor to $P(z)$ and is otherwise completely decoupled from the dynamics. In Figure \ref{fig:gr_sweep}, first-order coherence is computed on the left, full-width half maximum (FWHM) bandwidth calculated by the mean-field theory is shown on the right. Also shown is the analytical result calculated using extendon theory. In the fast gain recovery limit, one finds that the bandwidth is proportional to the extendon theory, as predicted by the adiabatic theory previously reported. However, for gain recovery times that are much longer than the coherence lifetime ($T_2$), one finds that the adiabatic theory begins to diverge, with the bandwidth narrowing and the ringing at the jump point eventually destabilizing. This can actually occur at timescales much shorter than the round-trip time, and is actually more closely related to $T_2$.

\section{Discussion}

The exact master equation and semi-exact mean-field theory in principle contain all of the dynamics of the Maxwell-Bloch equations, but do not have one of its limitations. One of the challenges associated with the Maxwell-Bloch equations is that they are in a sense an all-or-nothing proposition: to include the dynamics of realistic systems one must simulate the entire dynamics of the field, including all spatially-varying population and coherences. For an N-level system this would require keeping track of $N(N+1)/2-1$ quantities. However, for realistic multi-level laser systems, it is practically impossible to tie all of the parameters needed for these models to experimentally-measurable factors. The result is that one ends up including a large number of unknown parameters, and the simulation space grows enormous.

However, the non-adiabatic mean-field theory addresses this problem by essentially decoupling the calculation of the system's microscopic dynamics---the dynamics of the gain medium---from the macroscopic dynamics of the field. With this formalism, the gain medium becomes a lumped element: it affects the field only through the small-signal gain ($g_0$), the saturation field squared ($P_s$), the lineshape operator ($\hat{L}_2$), and the gain dynamics operator ($\hat{L}_1$). All of these quantities are directly measurable: gain and lineshape can be measured through single-pass gain measurements, the saturation field through output power measurements, and the gain dynamics through pump-probe measurements. The modeling of different systems can come just by changing these parameters. Furthermore, since $\hat{L}_1$ and $\hat{L}_2$ are readily computed in the frequency domain, this can be done by altering these functions. For example, inhomogeneous broadening can be incorporated by convolving $L_2(\omega)$ with a Gaussian. Similarly, heterogenous QCLs with multiple stacks could be modified by adding several Lorentzians together into $L_2(\omega)$. Off-resonant injection can be incorporated by shifting the center frequency of $L_2(\omega)$ (see Appendix \ref{offres}).

By the same token, non-trivial gain dynamics can be naturally incorporated into this formalism through the modification of $L_1(\omega)$. Though the adiabatic theory breaks down at long gain recovery times, it is possible to recover some of their dynamics using multi-component models that allow for the presence of both fast and slow components. For example, it is the case that the population inversion of a two-level laser can be efficiently described by the response function
\begin{align}
	\Delta n = \frac{1}{\frac{1}{T_1}+i \omega} P = T_1 L_1(\omega) P
\end{align}
where P is the pump rate. By contrast, a three-level system that has lasing between states 2 and 1 would have a response given by
\begin{align}
	\Delta n = \left(\frac{1}{\frac{1}{\tau_{21}}+i \omega} - \frac{1}{\frac{1}{\tau_{32}}+i \omega}\right)P,
\end{align}
provided that $\tau_{21} \gg \tau_{32}$. This is essentially a weighted sum of different $L_1(\omega)$'s. While the most accurate description of $L_1$ would actually be found through a solution to the Bloch equations, an adequate description that includes multiple components can be constructed as a weighted sum:
\begin{align}
	L_1(\omega) = \sum_i w_i L_{1i}(\omega),
\end{align}
where $L_{1i}(\omega) \equiv (1+i\omega T_{1i})^{-1}$. Note that within this model, one should take the $T_1$ that appears in $g_0$ and $P_s$ to be an effective value:
\begin{align}
	T_{1,\textrm{eff}} = \left( \sum_i \frac{w_i}{T_{1i}} \right)^{-1}.
\end{align}
Of course, the effective time constant does not drastically affect the dynamics for the reasons stated before.

\begin{figure}
	\includegraphics[width=\linewidth]{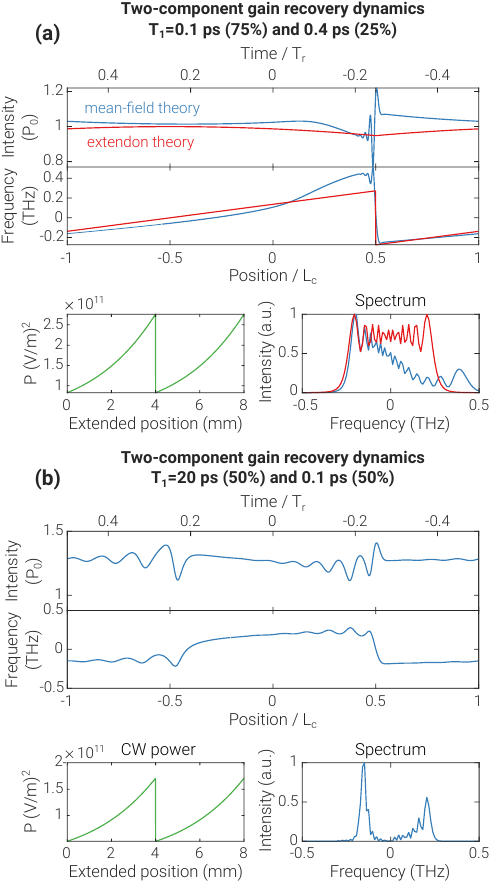}
	\caption{\label{fig:multi_comp}Two-component gain recovery dynamics with a dispersion of -1000 fs$^2$/mm. (a) Gain dynamics with a mixture of two fast components (0.1 ps and 0.4 ps), which is largely FM but has a nonlinear chirp. Also shown is the extendon theory, calculated using the effective gain recovery time. (b) Dynamics of a comb consisting of a slower gain component (20 ps) with a fast gain component (0.1 ps), which could be more descriptive of a medium like a terahertz QCL. While this can still be considered to be FM in the general sense that the intensity remains nonzero, it is more boxcar-like and has a bifurcated spectrum.}
\end{figure}

Figure \ref{fig:multi_comp} shows an example of multi-component dynamics for faster gain media and slower gain media, with Fig. \ref{fig:multi_comp}a illustrating an FM comb formed in a gain medium comprised of both 0.1 ps and 0.4 ps gain dynamics, while Fig. \ref{fig:multi_comp} shows an example of a gain medium comprised of 20 ps and 0.1 ps gain dynamics. Generally speaking, these models do not predict simple linear chirps, even if the spectrum is still FM in character. In the former case, one finds that one can still form FM combs that are smoothly chirped, but the chirping is considerably more nonlinear than in the single-component case, a consequence of different parts of the spectrum effectively experiencing different chirp rates. In the latter case, the chirp is boxcar-like and resembles the sorts of dynamics seen in double-peaked gain media, with a continuous spectrum that is strongly peaked at the edges. In both cases, stable combs are formed, and in the latter case stable combs can even form with zero dispersion (in contrast to extendons, which require a non-zero effective dispersion to be stable).

While in this work we have primarily focused on bidirectional cavities, the results discussed also have implications for comb formation in largely unidirectional cavities. For example, of late there has been interest in solitons formed by off-resonant injection Nozaki-Bekki solitons \cite{opacakNozakiBekkiSolitons2024}, other types of solitons structures in ring cavities \cite{mengDissipativeKerrSolitons2022,letsouLasingHybridizedSoliton2024}, and quantum walk combs \cite{heckelmannQuantumWalkLaser2023}. The analytical solutions that have been found usually require an adiabatic assumption, and in fact usually have an even more burdensome requirement of \textit{infinitely} fast gain. But at the same time, there has been numerical work revolving around the Maxwell-Bloch equations that has demonstrated that backscattering plays an important role in formation and stabilization in ring cavities \cite{seitnerBackscatteringInducedDissipativeSolitons2024}. Our theory can bridge these two gaps, allowing for the backscattering-induced contribution to be analytically evaluated, ultimately allowing for the fundamental limits to be discovered. It can also be used to analytically examine the role of non-adiabatic contributions to existing known soliton states, for example by using perturbation theory.

Finally, we also point out that there has also been significant interest in creating chip-scale versions of solid-state platforms like Ti:Sapphire \cite{yangTitaniumSapphireinsulatorIntegrated2024,wangPhotoniccircuitintegratedTitaniumSapphire2023}, which offer enormous gain bandwidths but usually also have long $T_1$'s ($\mu$s-level) to compensate for their very short $T_2$'s. While conventional Kerr lens mode-locking is not straightforward in these systems, it is feasible that other comb states relying on a fast component to the gain remain possible, in which case the theory presented here would be needed.

\section{Conclusion}
In conclusion, we have developed an exact master equation and a semi-exact mean-field theory to describe the dynamics of bidirectional active cavities without resorting to the adiabatic approximation. By introducing Lorentzian operators, we have created a comprehensive description that generalizes both fast and slow gain dynamics. Our operator-based formalism extends the applicability of mean-field theory to active cavities with significant internal gain and loss, as well as cavities with backscattering due to a gain grating. The resulting theory not only generalizes previous adiabatic models but also offer numerical stability and efficiency, facilitating simulations over a broad range of parameters. Importantly, our approach decouples the microscopic gain dynamics from the macroscopic field evolution, allowing for straightforward incorporation of complex gain media characteristics, including non-trivial lineshapes and multi-component gain recovery processes. Our findings are general and hold for any bidirectional or unidirectional active cavity, thereby extending the theoretical framework for a wide array of integrated lasers. This work paves the way for the discovery of novel comb states and could aid in the development of combs based on integrated platforms.

\section{Acknowledgments}

D.B. acknowledges Alex Dikopoltsev at ETH Zurich for valuable discussions, as well as support from from AFOSR grant no. FA9550-24-1-0349, ONR grant N00014-21-1-2735, AFOSR grant no. FA9550-20-1-0192, and NSF grant ECCS-2046772; this research is funded in part by the Gordon and Betty Moore Foundation through Grant GBMF11446 to the University of Texas at Austin to support the work of D.B.

\bibliography{burghoff_zotero}
\clearpage

\appendix
\begingroup
\large  
\onecolumngrid  

\renewcommand{\thesection}{\Alph{section}}  
\renewcommand{\thesubsection}{\thesection.\arabic{subsection}}  

\renewcommand{\section}[1]{%
	\refstepcounter{section}
	\setcounter{subsection}{0} 
	\par\bigskip\noindent
	\large\bfseries Appendix \thesection: #1\par\normalfont\bigskip
}

\renewcommand{\subsection}[1]{%
	\refstepcounter{subsection}
	\par\medskip\noindent
	\large\bfseries\thesubsection\quad #1\par\normalfont\medskip
}

\tcbset{highlight math style={colframe={rgb:red,55;green,126;blue,184}, colback=white,arc=4pt,boxrule=1pt}}

\section{Exact nonlinear polarization} \label{exactpol}

Our derivation is based on a first principles description of two-level systems. To find the nonlinear polarization of a two-level system, we begin with a derivation of the Bloch equations to establish our conventions. First, we assume that the density matrix is governed by the unperturbed Liouville equation as
\[
\dot{\rho} = \frac{1}{i \hbar} \left[ H, \rho \right] + \Gamma \rho
\]
where the Hamiltonian \( H \) is represented by the diagonal matrix
\[
H = \begin{pmatrix} E_g & 0 \\ 0 & E_e \end{pmatrix} \rightarrow \dot{\rho} = \begin{pmatrix} 0 & i\omega_0 \rho_{ge} \\ -i\omega_0 \rho_{eg} & 0 \end{pmatrix}.
\]
where \( \hbar \omega_0 = E_e - E_g \). In addition to this, we introduce the relaxation superoperator \( \Gamma \rho \), which accounts for decay and dephasing in the system:
\[
\Gamma \rho = \begin{pmatrix} -\Gamma_g \rho_{gg} + \Gamma_e \rho_{ee} & -\rho_{ge}/T_2 \\ -\rho_{eg}/T_2 & -\Gamma_e \rho_{ee} + \Gamma_g \rho_{gg} \end{pmatrix}
\]
where $T_2$ is the coherence lifetime. In the case of a two-level system, there are four parameters in the density matrix, but only two of them are independent. The following constraints hold for density matrices
\begin{align*}
\rho_{ee} + \rho_{gg} = 1 \\
\rho_{eg} = \rho_{ge}^*
\end{align*}
We define the coherence \( d \) and the population inversion $w$ as the off-diagonal element of the density matrix and the difference in excited and ground state populations, respectively:
\begin{align*}
	d &= \rho_{ge} \\
	w &\equiv \rho_{ee} - \rho_{gg}
\end{align*}
These definitions will be useful in analyzing the dynamics of the two-level system under various conditions. They evolve according to
\[
\dot{w} = \dot{\rho}_{ee} - \dot{\rho}_{gg} = - \Gamma_e \rho_{ee} + \Gamma_g \rho_{gg} - \Gamma_e \rho_{ee} + \Gamma_g \rho_{gg} 
= 2 \left( \Gamma_g \rho_{gg} - \Gamma_e \rho_{ee} \right)
\]
Solving $w=\rho_{ee} -\rho_{gg}$ and $1=\rho_{ee}+\rho_{gg}$ gives \( \rho_{gg} = \frac{1 - w}{2} \) and \( \rho_{ee} = \frac{1 + w}{2} \), and we get:
\[
\dot{w} = 2 \left( \Gamma_g \frac{1 - w}{2} - \Gamma_e \frac{1 + w}{2} \right) = -w (\Gamma_g + \Gamma_e) + (\Gamma_g - \Gamma_e)
\]
Thus:
\[
\dot{w} = \left( -w - \frac{\Gamma_g - \Gamma_e}{\Gamma_g + \Gamma_e} \right) \left( \Gamma_g + \Gamma_e \right)
= -\frac{w - w_0}{T_1}
\]
where \( w_0 = \frac{\Gamma_g - \Gamma_e}{\Gamma_g + \Gamma_e} \) is the steady-state solution (positive only for \( \Gamma_g > \Gamma_e \)) and $\frac{1}{T_1} \equiv \Gamma_g + \Gamma_e$
is the inversion decay rate (in terms of the population lifetime $T_1$). For the coherence, we have
\[
\dot{d} = i \omega_0 \rho_{ge} - \frac{\rho_{ge}}{T_2} = \left( i \omega_0 - \frac{1}{T_2} \right) d
\]
whose solution rotates at \( \omega_0 \) according to $d \sim e^{i \omega_0 t} e^{-t/T_2}$. While \( w \) decays to steady state, \( d \) oscillates and decays.

Assume the electric field takes the form \( E_f(t) = E_a(t) + E_a^*(t) \), where \( E_a = A(t)e^{i\omega_0 t} \), and $A$ is the slowly varying envelope. (In this convention, the intensity would be found from the index $n_0$ as $I = 2 n_0 c \varepsilon_0 |E_a|^2$.) Our task is to find the polarization density $P$ induced by this field. The perturbation Hamiltonian is
\[
H' = -e E_f x = \begin{pmatrix} 0 & \mu \\ \mu^* & 0 \end{pmatrix} E_f
\]
The total time evolution of the density matrix is given by
\[
\dot{\rho} = \frac{1}{i\hbar} \left[ H_0 + H', \rho \right] + \Gamma \rho
\]
The commutator \( [H', \rho] \) becomes
\begin{align*}
	[H', \rho] &= E_f(t) \left[ \begin{pmatrix} 0 & \mu \\ \mu^* & 0 \end{pmatrix} \begin{pmatrix} \rho_{gg} & \rho_{ge} \\ \rho_{eg} & \rho_{ee} \end{pmatrix} - \begin{pmatrix} \rho_{gg} & \rho_{ge} \\ \rho_{eg} & \rho_{ee} \end{pmatrix} \begin{pmatrix} 0 & \mu \\ \mu^* & 0 \end{pmatrix} \right] \\
	&= E_f(t) \left( \begin{pmatrix} \rho_{eg} \mu^* & \rho_{ee} \mu \\ \rho_{gg} \mu^* & \rho_{ge} \mu \end{pmatrix} - \begin{pmatrix} \rho_{ge} \mu & \rho_{gg} \mu \\ \rho_{ee} \mu^* & \rho_{eg} \mu^* \end{pmatrix} \right) \\
	&= E_f(t) \begin{pmatrix} d^* \mu^* - d \mu & w \mu \\ -w \mu^* & d \mu - d^* \mu^* \end{pmatrix}
\end{align*}
Thus, the equations for the evolution of \( w \) and \( d \) are
\begin{align}
	\dot{w} &= \frac{1}{i\hbar} 2 \left( d\mu^* - d^*\mu \right) E_f(t) - \frac{w - w_0}{T_1} \nonumber \\
	\dot{d} &= \frac{1}{i\hbar} \left( w \mu \right) E_f(t) + \left( i \omega_0 - \frac{1}{T_2} \right) d \label{wddynamics} 
\end{align}

Next, we apply the rotating wave approximation and assume that the coherence approximately rotates at $\omega_0$ as the population inversion does not. Defining demodulated parameters as \( d(t) \equiv \tilde{d}(t) e^{i\omega_0 t} \) and \( w(t) \equiv \tilde{w}(t) e^{i 0 t} = \tilde{w}(t)\) (which are assumed to vary slowly in time), we have
\begin{align*}
	\dot{w} &= \dot{\tilde{w}} = \frac{1}{i\hbar} 2 \left( \tilde{d} e^{i\omega_0 t} \mu^* - \tilde{d}^* e^{-i\omega_0 t} \mu \right) \left( A e^{i\omega_0 t} + A^* e^{-i\omega_0 t} \right) - \frac{w - w_0}{T_1}
	\\
	\dot{d} &=  \dot{\tilde{d}} e^{i\omega_0 t} + \tilde{d} \left( i\omega_0 \right) e^{i\omega_0 t} = \frac{1}{i\hbar} \left( \tilde{w} \mu \right) \left( A e^{i\omega_0 t} + A^* e^{-i\omega_0 t} \right) 
	+ \left( i\omega_0 - \frac{1}{T_2} \right) \tilde{d} e^{i\omega_0 t}
\end{align*}
which gives
\begin{align*}
	\dot{\tilde{w}} &= \frac{2}{i\hbar}  \left( \tilde{d}A\mu^* e^{i 2\omega_0 t} + \tilde{d}A^*\mu - \tilde{d}^*  \mu A - \tilde{d}^* \mu A^* e^{-i 2\omega_0 t} \right) - \frac{w - w_0}{T_1}
	\\
	\dot{\tilde{d}} &= \frac{1}{i\hbar} \tilde{w} \mu \left( A + A^* e^{-i2\omega_0 t} \right) - \frac{\tilde{d}}{T_2}.
\end{align*}
Neglecting fast terms at \( \pm 2\omega_0 \), the evolution equations become
\begin{align}
	\dot{\tilde{w}} &= \frac{2}{i\hbar} \left( \tilde{d}^* \mu^* A - \tilde{d} \mu A^* \right) - \frac{\tilde{w} - w_0}{T_1} \label{wtildedot} \\
	\dot{\tilde{d}} &= \frac{1}{i\hbar} \tilde{w} \mu A - \frac{\tilde{d}}{T_2}. \label{dtildedot}
\end{align}
Solving the equation for \( \dot{\tilde{d}} \) yields
\begin{alignat*}{2}
	\left( \frac{1}{T_2} + \frac{\partial}{\partial t} \right) \tilde{d} &= \frac{1}{i\hbar} \tilde{w} \mu A
\end{alignat*}
or
\begin{alignat}{2}
	\tilde{d} &= \left( \frac{1}{T_2} + \frac{\partial}{\partial t} \right)^{-1} \frac{1}{i\hbar} \tilde{w} \mu A \nonumber \\
	&= \frac{T_2}{i\hbar} \mu \hat{L}_2 \tilde{w} A \label{dsoln}
\end{alignat}
where we have defined the Lorentzian operator \( \hat{L}_i \) as
\begin{alignat*}{2}
	\hat{L}_i f &\equiv \left( 1 + T_i \frac{\partial}{\partial t} \right)^{-1}f =  \int_{-\infty}^{t} dt' f(t') \frac{1}{T_i} e^{-(t - t')/T_i}.
\end{alignat*}
The integral identity is proven in Appendix \ref{lorfactor}; it allows this operator to be expressed as a convolution with an impulse response. Importantly, \( \hat{L}_i \) does not commute with multiplication. Inserting (\ref{dsoln}) into (\ref{wtildedot}) gives
\begin{alignat*}{2}
	&\hspace{4mm}\left( \frac{1}{T_1} + \frac{\partial}{\partial t} \right) \tilde{w} 
	= \frac{2}{i\hbar} \left( \tilde{d} \mu^* A^* - \tilde{d}^* \mu A \right) + \frac{w_0}{T_1}\\
	\tilde{w} &= \left( \frac{1}{T_1} + \frac{\partial}{\partial t} \right)^{-1} \frac{2}{i\hbar} \left( \tilde{d} \mu^* A^* - \tilde{d}^* \mu A \right) 
	+ \left( \frac{1}{T_1} + \frac{\partial}{\partial t} \right)^{-1} \frac{w_0}{T_1} \\
	&= T_1 \hat{L}_1 \left( \frac{2}{i\hbar} \left( \tilde{d} \mu^* A^* - \tilde{d}^* \mu A \right) \right) 
	+ T_1 \hat{L}_1 w_0
\end{alignat*}
Since $w_0$ is not time-dependent, \( \hat{L}_1 w_0 = w_0 \). We can now express \( \tilde{w} \) as:
\begin{alignat*}{2}
	\tilde{w} &= \frac{2 T_1}{i\hbar} \hat{L}_1 \left[ \mu^* A^*  \frac{T_2}{i\hbar} \mu \hat{L}_2 \tilde{w} A - \mu A \frac{T_2}{-i\hbar} \mu^* \left( \hat{L}_2 \tilde{w} A \right)^* \right] + w_0 \\
	&= -\frac{2 T_1 T_2}{\hbar^2} |\mu|^2 \hat{L}_1 \left[ A^* \hat{L}_2 \tilde{w} A + A  \hat{L}_2^* A^* \tilde{w} \right] + w_0 \\
	&= -\frac{2 T_1 T_2}{\hbar^2} |\mu|^2 \hat{L}_1 \left( A^* \hat{L}_2 A + A  \hat{L}_2^* A^* \right) \tilde{w} + w_0 \\
	\tilde{w} &= \left[ 1 + \frac{2 T_1 T_2}{\hbar^2} |\mu|^2 \hat{L}_1 \left( A^* \hat{L}_2 A + A  \hat{L}_2^* A^* \right) \right]^{-1} w_0.
\end{alignat*}
Inserting this back into (\ref{dsoln}), we obtain:
\begin{alignat*}{2}
	\tilde{d} &= \frac{T_2}{i \hbar} \mu \hat{L}_2 A 
	\left[ 1 + \frac{2 T_1 T_2}{\hbar^2} | \mu |^2 \hat{L}_1 \left( A^* \hat{L}_2 A + A \hat{L}_2^* A^* \right) \right]^{-1} w_0 \\
	&= \frac{w_0 T_2 \mu}{i \hbar} \hat{L}_2 A
	\left[ 1 + \frac{2 T_1 T_2}{\hbar^2} | \mu |^2 \hat{L}_1 \left( A^* \hat{L}_2 A + A \hat{L}_2^* A^* \right) \right]^{-1}.
\end{alignat*}
Critically, while the bracketed term was originally an operator, it has now been converted into a scalar quantity, as it was operating on the constant $w_0$. When \( \frac{\partial}{\partial t} \approx 0 \) and $\hat{L}_2 = \hat{I}$ , it simplifies to the familiar intensity saturation:
\begin{alignat*}{2}
	1 + \frac{2 T_1 T_2}{\hbar^2} |A|^2 (2|A|^2) &= 1 + \frac{|A|^2}{P_s}
\end{alignat*}
where $P_s \equiv E_s^2 \equiv \left( \frac{4 T_1 T_2}{\hbar^2} | \mu |^2 \right)^{-1}$ is the saturation field squared. Note that saturation field squared is proportional to intensity.

Now, we can calculate the polarization density \( P \) as
\begin{align}
	P &= -e N \langle x \rangle = -e N \text{Tr} \left( \rho x \right) \nonumber \\
	&= - N \text{Tr} \left( \begin{pmatrix} \rho_{gg} & \rho_{ge} \\ \rho_{eg} & \rho_{ee} \end{pmatrix} \begin{pmatrix} 0 & \mu \\ \mu^* & 0 \end{pmatrix} \right) = - N \left( \rho_{ge} \mu^* + \rho_{eg} \mu \right) \nonumber \\
	&= - N \left( d \mu^* + d^* \mu \right).
\end{align}
where $N$ is the doping, $e$ is the electron charge magnitude, and $\mu \equiv e \langle g|\hat{x}|e\rangle$ is the dipole moment. The polarization density \( P \) can then be found to be:
\begin{alignat*}{2}
	P &= - N \left( \mu^* d + \mu d^* \right) 
	= -N \left( e^{i \omega_0 t} \mu^* \tilde{d} + e^{-i \omega_0 t} \mu \tilde{d}^* \right) \\
	&= \frac{N |\mu|^2 w_0 T_2}{\hbar} \left( i e^{i \omega_0 t} \hat{L}_2 A 
	\left[ 1 + \frac{2 T_1 T_2}{\hbar^2} | \mu |^2 \hat{L}_1 \left( A^* \hat{L}_2 A + A \hat{L}_2^* A^* \right) \right]^{-1} + \text{c.c.} \right).
\end{alignat*}
Finally, we can express P in terms of its positive frequency component as \( P = P(\omega_0) e^{i \omega_0 t} + P^*(\omega_0)e^{-i \omega_0 t}\), where
\begin{empheq}[box=\tcbhighmath]{align}
	P(\omega_0) &= i \frac{N |\mu|^2 w_0 T_2}{\hbar}   \hat{L}_2 A 
	\left[ 1 + \frac{1}{2 P_s} \hat{L}_1 \left( A^* \hat{L}_2 A + A (\hat{L}_2 A)^* \right) \right]^{-1}
\end{empheq}
with $\hat{L}_i \equiv \left( 1 + T_i \frac{\partial}{\partial t} \right)^{-1}$ reflects the  lineshape (in the case of $\hat{L}_2$) and the population dynamics (in the case of $\hat{L}_1$). Note that the prefactor of $\hat{L}_2$ applies to both the product of A and the inverted term.

\section{Exact master equations in a bidirectional cavity} \label{exactmaster}
For envelopes corresponding to two counterpropagating waves, the total electric field envelope $A(z,t)$ is a sum of waves traveling in each direction:
\begin{alignat*}{2}
	A(z,t) &= E_+(z,t) e^{-i k_0 z} + E_-(z,t) e^{i k_0 z} \\
	E_f(z,t) &= \left( E_+ e^{-i k_0 z} + E_- e^{i k_0 z} \right) e^{i \omega_0 t} + \text{c.c.}
\end{alignat*}
where \( E_+ \) and \( E_- \) are the slowly varying envelopes of the forward and backward propagating waves, respectively. The nonlinear wave equation for each envelope is
\begin{equation}
	\frac{1}{v_g} \frac{\partial E_\pm}{\partial t} \pm \frac{\partial E_\pm}{\partial z} = i \frac{1}{2 \omega_0 n_0 \varepsilon_0 c} e^{-i \left( \omega_0 t \mp k_0 z \right)} \frac{\partial^2 }{\partial t^2} P
	\equiv e^{\pm i k_0 z} \nlgainvar_\text{NL} \label{nlwaveeqn}
\end{equation}
where $v_g$ is the group velocity and we have introduced the nonlinear gain term \( \nlgainvar_\text{NL} \) as 
\[
\nlgainvar_\text{NL} = i \frac{1}{2 \omega_0 n_0 \varepsilon_0 c} e^{-i \omega_0 t}\frac{\partial^2 }{\partial t^2} P
\approx \frac{i}{2 n_0 \varepsilon_0 c} e^{-i \omega_0 t} (i \omega_0)^2 P(\omega_0). 
\]
In the above expression, we take only the positive frequency component of the nonlinear polarization and use the slowly-varying envelope approximation when performing the differentiation. Combining this with the previous result yields
\begin{align*}
	\nlgainvar_\text{NL} &= -i \frac{\omega_0}{2 n_0 \varepsilon_0 c} i \frac{N  \left| \mu \right|^2  w_0  T_2}{\hbar} \hat{L}_2 \left[ 1 + \frac{1}{P_s} \hat{L}_1 \operatorname{Re} \left( A^* \hat{L}_2 A \right) \right]^{-1} A \\
	&= \frac{N \left| \mu \right|^2 w_0 \omega_0 T_2}{2 n_0 \varepsilon_0 c \hbar} \hat{L}_2 \left[ 1 + \frac{1}{P_s} \hat{L}_1 \operatorname{Re} \left( A^* \hat{L}_2 A \right) \right]^{-1} A
\end{align*}
where the units of the prefactor are in \( \text{cm}^{-1} \). We will therefore define
\[
g_0 = \frac{N w_0 \left| \mu \right|^2  \omega_0 T_2}{n_0 \varepsilon_0 c \hbar}
\]
as the small-signal power gain of the medium. (For a quantum cascade laser with an current density $J$, the population inversion density is approximately $N w_0 = \frac{J T_1}{e L_{mod}}$, although the details depend on the specifics of the gain medium.) We can therefore simplify the nonlinear term as
\begin{equation}
	\nlgainvar_\text{NL} = \frac{g_0}{2} \hat{L}_2 \left[ 1 + \frac{1}{P_s} \hat{L}_1 \operatorname{Re} \left( A^* \hat{L}_2 A \right) \right]^{-1} A \equiv \frac{\hat{g}_{NL}}{2} A \label{fNL}
\end{equation}
where we have introduced the nonlinear gain operator and once again \( \hat{L}_i = \left( 1 + T_i \frac{\partial}{\partial t} \right)^{-1} \). In the adiabatic limit, the nonlinear term recovers the standard saturated gain, since:

\begin{enumerate}
	\item The \( \hat{L}_2 \) term reflects the Lorentzian lineshape. For \( \frac{\partial}{\partial t} = i\omega \), one has
	\[
	\left( 1 + T_2 \frac{\partial}{\partial t} \right)^{-1} A = \frac{1}{1 + i\omega T_2} A = \frac{1 - i \omega T_2}{1 + \left( \omega T_2 \right)^2} A
	\]
	where \( L_2(\omega) \equiv \frac{1}{1 + i\omega T_2} \) is a complex Lorentzian factor, whose real part is \( \left( 1 + \left( \omega T_2 \right)^2 \right)^{-1} \). It has a full-width half-maximum (FWHM) of
	\[
	\omega_\text{HWHM} = \frac{1}{T_2}, \quad f_\text{FWHM} = \frac{1}{\pi T_2}.
	\]
	
	\item In steady state, when \( \hat{L}_1 = \hat{I}\), this reduces to:
	\[
	\nlgainvar_\text{NL} = \frac{g_0}{2} L(\omega) A \frac{1}{1 + \frac{1}{P_s} |A|^2 L(\omega)}
	\]
	which is the usual form of the saturated gain.
\end{enumerate}
To find the effect of counterpropagating waves on each other, we note that the $e^{\pm i k_0 z}$ term on the right hand side of the nonlinear wave equation (\ref{nlwaveeqn}) will select only the spatial Fourier component at $e^{\mp i k_0 z}$. We will therefore introduce the positive and negative spatial frequency components of $\nlgainvar_\text{NL}$ as $\nlgainvar_\text{NL} = \nlgainvar_\text{NL}^+ e^{-i k_0 z} + \nlgainvar_\text{NL}^- e^{+i k_0 z}$, in which case
\begin{align}
	\frac{1}{v_g} \frac{\partial E_\pm}{\partial t} \pm \frac{\partial E_\pm}{\partial z} = \nlgainvar_\text{NL}^\pm
\end{align}
and
\begin{align*}
	\nlgainvar_\text{NL}^\pm &= \left( \frac{2\pi}{k_0} \right)^{-1} \int_0^{2\pi/k_0} dz \, e^{\pm i k_0 z} \nlgainvar_\text{NL} \left( A \to E_+ e^{-i k_0 z} + E_- e^{+i k_0 z} \right)
\end{align*}
Inserting (\ref{fNL}) and exploiting the fact that \( e^{\pm i k_0 z} \) has no temporal dependence and therefore commutes with \( \hat{L}_i \), we can write
\begin{alignat*}{2}
	\nlgainvar_\text{NL}^\pm = \left( \frac{2 \pi}{k_0} \right)^{-1} \int_0^{2\pi/k_0} dz \, &e^{\pm i k_0 z} \frac{g_0}{2} \left( e^{-i k_0 z} \hat{L}_2 E_+ + e^{+i k_0 z} \hat{L}_2 E_- \right) \\
	\times \Bigg[ 1 + \frac{1}{2 P_s} &\Big( \hat{L}_1 E_+^* \hat{L}_2 E_+ + e^{+i 2 k_0 z} \hat{L}_1 E_+^* \hat{L}_2 E_- \\
	&+ e^{-i 2 k_0 z} \hat{L}_1 E_-^* \hat{L}_2 E_+ + \hat{L}_1 E_-^* \hat{L}_2 E_- + \text{c.c.} \Big) \Bigg]^{-1}.
\end{alignat*}
We can expand this out by introducing the saturation-like terms \( D_0 \) and \( D_2^{\pm} \):
\begin{alignat*}{2}
	D_0 &\equiv 1 + \frac{1}{P_s} \operatorname{Re} \hat{L}_1 \left( E_+^* \hat{L}_2 E_+ + E_-^* \hat{L}_2 E_- \right) > 1 \\
	D_2^{\pm} &\equiv \frac{1}{P_s} \hat{L}_1 \left( E_\pm \left(\hat{L}_2 E_\mp\right)^* + E_\mp^* \hat{L}_2 E_\pm \right)
\end{alignat*}
which give:
\begin{alignat*}{2}
	\nlgainvar_\text{NL}^\pm &= \left( \frac{2\pi}{k_0} \right)^{-1} \int_0^{2\pi/k_0} dz \, e^{\pm i k_0 z} \frac{g_0}{2} 
	\left( e^{\mp i k_0 z} \hat{L}_2 E_\pm + e^{\pm i k_0 z} \hat{L}_2 E_\mp \right) \\
	&\hspace{2cm}\times \Bigg[ 1 + \frac{1}{2 P_s} \Big( \hat{L}_1 E_\pm^* \hat{L}_2 E_\pm + e^{\pm i 2 k_0 z} \hat{L}_1 E_\pm^* \hat{L}_2 E_\mp \\
	&\hspace{2.5cm}+ e^{\mp i 2 k_0 z} \hat{L}_1 E_\mp^* \hat{L}_2 E_\pm + \hat{L}_1 E_\mp^* \hat{L}_2 E_\mp + \text{c.c.} \Big) \Bigg]^{-1} \\
	&= \left( \frac{2 \pi}{k_0} \right)^{-1} \int_0^{2\pi/k_0} dz \, \frac{g_0}{2} \left(\hat{L}_2 E_\pm + e^{\pm i 2 k_0 z} \hat{L}_2 E_\mp \right) \\
	&\hspace{2cm}\times \Bigg[ 1 + D_0 - 1 + e^{\mp i 2 k_0 z} \frac{1}{2} D_2^{\pm} + e^{\pm i 2 k_0 z} \frac{1}{2} D_2^{\pm *} \Bigg]^{-1} \\
	&= \left( \frac{2 \pi}{k_0} \right)^{-1} \frac{g_0}{2} \Bigg( \hat{L}_2 E_\pm \int_0^{2\pi/k_0} \left( D_0 + |D_2^\pm| \cos(2k_0 z + \angle D_2^\pm) \right)^{-1} dz \\
	&\hspace{2.1cm}+ \hat{L}_2 E_\mp \int_0^{2\pi/k_0} \left( D_0 + |D_2^\pm| \cos(2k_0 z + \angle D_2^\pm) \right)^{-1} e^{\pm i 2k_0 z} dz \Bigg) \\
\end{alignat*}
Changing variable to $2 k_0 z + \angle D_2^\pm \equiv 2k_0 z'$, we have $z' = z + \frac{\angle D_2^\pm}{2k_0}$ and $dz' = dz$. The integration limits can be unchanged due to periodicity, and defining \mbox{\( x\equiv |D_2^\pm|/D_0<1 \)},
\begin{alignat*}{2}	
	\nlgainvar_\text{NL}^\pm &= \left( \frac{2 \pi}{k_0} \right)^{-1} \frac{g_0}{2} \Bigg( \hat{L}_2 E_\pm D_0^{-1} \int_0^{2\pi/k_0} \left( 1 + x \cos(2k_0 z) \right)^{-1} dz \\
	&\hspace{2.1cm}+ \hat{L}_2 E_\mp D_0^{-1} \int_0^{2\pi/k_0} \left( 1 + x \cos(2k_0 z) \right)^{-1} e^{\pm i (2k_0 z - \angle D_2^\pm)} dz \Bigg) \\
	&= \left( \frac{2 \pi}{k_0} \right)^{-1} \frac{g_0}{2} \Bigg( \hat{L}_2 E_\pm D_0^{-1} \frac{2}{2k_0} \frac{2\pi}{\scalebox{0.8}{$\sqrt{1 - |D_2^\pm|^2/D_0^2}$}}  \\
	&\hspace{2.1cm}+ \hat{L}_2 E_\mp D_0^{-1} \frac{2}{2k_0}   \left( 2\pi - \frac{2\pi}{\scalebox{0.8}{$\sqrt{1 - |D_2^\pm|^2/D_0^2}$}} \right) \frac{D_0}{|D_2^\pm|} e^{\mp i \angle D_2^\pm} \Bigg) \\
	&= \frac{g_0}{2} \hat{L}_2 \frac{1}{\sqrt{D_0^2 - |D_2^\pm|^2}} \left(  E_\pm  +  E_\mp  \left( \sqrt{D_0^2 - |D_2^\pm|^2} - D_0 \right) \frac{1}{D_2^{\pm *}} \right).
\end{alignat*}
Finally, we have obtained an exact master equation description that describes the interaction of counterpropagating waves via a gain medium:
\begin{empheq}[box=\tcbhighmath]{align}
	&\frac{1}{v_g} \frac{\partial E_\pm}{\partial t} \pm \frac{\partial E_\pm}{\partial z} = \nlgainvar_\text{NL}^\pm \nonumber \\
	&\nlgainvar_\text{NL}^\pm = \frac{g_0}{2} \hat{L}_2  \frac{1}{\scalebox{1}{$\sqrt{D_0^2 - |D_2|^2}$}} \left( E_\pm + E_\mp \frac{1}{D_2^{\pm *}} \left(\scalebox{1}{$\sqrt{D_0^2 - |D_2|^2} - D_0$} \right) \right) \\
	&\quad D_0 \equiv 1 + \frac{1}{P_s} \operatorname{Re} \hat{L}_1 \left( E_+^* \hat{L}_2 E_+ + E_-^* \hat{L}_2 E_- \right) \nonumber \\
	&\quad D_2^\pm \equiv \frac{1}{P_s} \hat{L}_1 \left( E_\pm (\hat{L}_2 E_\mp)^* + E_\mp^* \hat{L}_2 E_\pm \right). \nonumber 
\end{empheq}
Note that \( D_2^{\pm *} = D_2^\mp \), so we have introduced \( |D_2|^2 \equiv |D_2^{\pm}|^2 = |D_2^\mp|^2 \) to emphasize their common prefactor. This formulation is for all intents and purposes exact, and clarifies the role of saturation, backscattering, and each Bloch lifetime.
\begin{enumerate}
	\item Conventional gain saturation appears as the $\left(D_0^2 - |D_2|^2\right)^{-1/2}$ prefactor.
	\item The saturation-like terms ($D_0$ and $D_2^\pm$) consist of a summation of multiple power-like components second order in the field. They are related to population and coherence, respectively.
	\item The $\hat{L}_2$ operator associated with dephasing applies to the whole expression (including gain saturation), but also applies to half of each power-like components. 
	\item The $\hat{L}_1$ operator associated with the gain recovery time applies to all of the power-like components, essentially low-pass filtering them.
	\item As we will show, the backscattering term associated with the counterpropagating wave $E_\mp$ is the main driver for the formation of FM combs.
	
\end{enumerate}
To see why backscattering alone is necessary for FM comb generation, consider the low-field limit, in which 
\begin{alignat*}{2}
	\sqrt{D_0^2 - |D_2|^2} &\approx D_0 \\
	\frac{1}{D_2^{\pm *}} \left( \sqrt{D_0^2 - |D_2|^2} - D_0 \right) &\approx - \frac{1}{2} D_2^{\pm}.
\end{alignat*}
\vspace{-5mm}
\begin{alignat*}{2}
	\nlgainvar_\text{NL}^\pm &\approx \frac{g_0}{2} \hat{L}_2 \frac{1}{D_0} \left( E_\pm - \frac{1}{2} E_\mp D_2^{\pm} \right) \\
	&= \frac{g_0}{2} \hat{L}_2 \frac{1}{1 + \frac{1}{P_s} \operatorname{Re} \hat{L}_1 \left( E_+^* \hat{L}_2 E_+ + E_-^* \hat{L}_2 E_- \right)} \\
	&\quad \times \left( E_\pm - \frac{1}{2 P_s} E_\mp \hat{L}_1 \left( E_\pm \left( \hat{L}_2 E_\mp \right)^* + E_\mp^* \hat{L}_2 E_\pm \right) \right)
\end{alignat*}
Without time variation, \( \hat{L}_1 = \hat{L}_2 = \hat{I} \), $D_0 = 1 + \frac{1}{P_s} \left( |E_+|^2 + |E_-|^2 \right)$, and $D_2^\pm = \frac{2}{P_s}  E_\pm E_\mp^*$, which would lead to
\begin{alignat*}{2}
	\nlgainvar_\text{NL}^\pm &\approx \frac{g_0}{2} \hat{L}_2 \frac{1}{1 + \frac{1}{P_s} \left( |E_+|^2 + |E_-|^2 \right)} 
	\left( E_\pm - \frac{1}{2} E_\mp \frac{2}{P_s}  E_\pm E_\mp^* \right) \\
	&= \frac{g_0}{2} \hat{L}_2 \frac{1}{1 + \frac{1}{P_s} \left( |E_+|^2 + |E_-|^2 \right)} 
	\left( 1 - \frac{1}{P_s} |E_\mp|^2 \right) E_\pm
\end{alignat*}
which is the usual form seen for gain saturation in the presence of backscattering. Next, if we allow for time variation and define the saturated gain as
\[
g_s \equiv \frac{g_0}{\sqrt{D_0^2 - |D_2^{\pm}|^2}}
\]
then in the low-field ($\mathcal{O}(E^3)$) and adiabatic ($\mathcal{O}(\frac{\partial}{\partial t})$) limits we find that
\begin{alignat*}{2}
	\nlgainvar_\text{NL}^\pm &\approx \frac{1}{2} \hat{L}_2 \, g_s \left( E_\pm - \frac{1}{2 P_s} E_\mp \hat{L}_1 \left( E_\pm \left( \hat{L}_2 E_\mp \right)^* + E_\mp^* \hat{L}_2 E_\pm \right) \right) \\
	&\approx \frac{1}{2} \hat{L}_2 \,  g_s \Big( E_\pm - \frac{1}{2 P_s} E_\mp \left( E_\pm \scalebox{1}{$\left( E_\mp - T_2 \frac{\partial E_\mp}{\partial z} \right)$}^* - T_1 \frac{\partial E_\pm}{\partial t} E_\mp^* - T_1 \frac{\partial E_\mp^*}{\partial t} E_\pm \right) \\
	&\hspace{3.7cm} + E_\mp^* \scalebox{1}{$\left( E_\pm - T_2 \frac{\partial E_\mp}{\partial z}\right)$} \hspace{1mm}- T_1 \frac{\partial E_\mp^*}{\partial t}E_\pm - T_1 \frac{\partial E_\pm}{\partial t} E_\mp^* \Big) \\
	&= \frac{1}{2} \hat{L}_2 \, g_s \Big( E_\pm - \frac{1}{2 P_s} E_\mp \scalebox{.93}{$\left( 2 E_\pm E_\mp^* - (2 T_1 + T_2) \frac{\partial E_\mp^*}{\partial t} E_\pm - (2 T_1 + T_2) \frac{\partial E_\pm}{\partial t} E_\mp^* \right)$} \Big) \\
	&= \frac{1}{2} \hat{L}_2 \, g_s \left( 1 - \frac{1}{P_s} |E_\mp|^2 + \scalebox{1}{$ \left( T_1 + \frac{1}{2} T_2 \right)$} \left( \frac{\partial E_\mp^*}{\partial t} E_\pm + |E_\mp|^2 \frac{\partial}{\partial t} \right) \right) E_\pm.
\end{alignat*}
The \( T_1 + \frac{1}{2} T_2 \) term appears in the phase potential and is FM-generating. Therefore, the backscattering term alone can be considered its origin. Expanding the whole expression in the adiabatic limit and to low-order (including $g_s$) yields
\begin{alignat*}{2}
	\nlgainvar_\text{NL}^\pm &= \frac{g_0}{2} \Bigg( 1 - T_2 \frac{\partial}{\partial t} - \frac{1}{P_s} \left( |E_\pm|^2 + 2 |E_\mp|^2 \right)  \\
	&\hspace{.8cm}+ \frac{1}{P_s} \Bigg( \left( T_1 + \tfrac{5}{2} T_2 \right) \left( |E_\mp|^2 \frac{\partial}{\partial t} + E_\mp^* \frac{\partial E_\mp}{\partial t} + |E_\pm|^2 \frac{\partial}{\partial t} \right) \\
	&\hspace{1.6cm}+ \left( 2 T_1 + 3 T_2 \right) \frac{\partial E_\mp^*}{\partial t} E_\mp + \left( T_1 + \tfrac{3}{2} T_2 \right) \frac{\partial E_\pm^*}{\partial t} E_\pm \Bigg) E_+,
\end{alignat*}
agreeing with the prior result.

\section{Cavity extension and power normalization} \label{extnorm}
The main difference between active cavities and passive cavities lies in the size of their internal gain and losses. In a passive cavity such as a microresonator comb, the losses and nonlinear gains are typically very small, on account of the inherent smallness of the optical nonlinearity. However, that is not the case in active cavities---a semiconductor Fabry-Perot laser, for example, typically has enormous changes in the field at each facet. This prevents the construction of a mean-field theory, as the core assumption of mean-field theory is that the changes are small. However, its assumptions can be recovered by noticing that the power dynamics of a laser above threshold are strikingly similar to their continuous wave (CW) dynamics. By normalizing to the CW power, the wave becomes periodic over a round trip and mean-field theory can be created.

The first order of business is to extend the cavity so that the bidirectional field is unified as a unidirectional one. To do this, we flip the negative propagating wave and glue it to the positive wave. If the physical cavity coordinates are at $z \in [0,L_c]$, then $E_-$ is placed at $z \in [-L_c,0]$. That is, we define
\[
E(z,t) \equiv
\begin{cases}
	E_+(z,t) & \text{for } z > 0 \\
	E_-(-z,t) & \text{for } z < 0
\end{cases}
\]
\[
\frac{\partial E}{\partial t} =
\begin{cases}
	\frac{\partial E_+}{\partial t}(z,t) & z \geq 0 \\
	\frac{\partial E_-}{\partial t}(-z,t) & z < 0
\end{cases}
\quad \quad
\frac{\partial E}{\partial z} =
\begin{cases}
	\frac{\partial E_+}{\partial z}(z,t) & z \geq 0 \\
	-\frac{\partial E_-}{\partial z}(-z,t) & z < 0
\end{cases}
\]
\[
\hat{L}_i E =
\begin{cases}
	(\hat{L}_i E_+)(z,t) & z \geq 0 \\
	(\hat{L}_i E_-)(-z,t) & z < 0
\end{cases}
\]
Evaluating at \( z > 0 \) and introducing $E_{+z}\equiv E(z,t)$ and $E_{-z}\equiv E(-z,t)$:
\begin{alignat*}{2}
	\frac{1}{v_g} \frac{\partial E}{\partial t} + \frac{\partial E}{\partial z} &= \frac{1}{v_g} \frac{\partial E_+}{\partial t}(z,t) - \frac{\partial E_+}{\partial z}(z,t) \\ &= \frac{g_0}{2} \hat{L}_2 \frac{1}{\sqrt{D_0^2 - |D_2|^2}} \left( E_+ + E_- \frac{1}{D_2^{+*}} \left( \scalebox{1}{$\sqrt{D_0^2 - |D_2|^2}$} - D_0 \right) \right) \\
	&= \frac{g_0}{2} \hat{L}_2 \frac{1}{\sqrt{D_0^2 - |D_2|^2}} \left( E_{+z} + E_{-z}\frac{1}{D_2^{+*}} \left( \scalebox{1}{$\sqrt{D_0^2 - |D_2|^2}$} - D_0 \right) \right) \\
	D_0 &= 1 + \frac{1}{P_s} \operatorname{Re} \hat{L}_1 \left( E_+^* \hat{L}_2 E_+ + E_-^* \hat{L}_2 E_- \right) \\
	&= 1 + \frac{1}{P_s} \operatorname{Re} \hat{L}_1 \left( E_{+z}^* \hat{L}_2 E_{+z} + E_{-z}^* \hat{L}_2 E_{-z} \right) \\
	D_2^+ &= \frac{1}{P_s} \hat{L}_1 \left( E_+ \left( \hat{L}_2 E_- \right)^* + E_-^* \hat{L}_2 E_+ \right) \\
	&= \frac{1}{P_s} \hat{L}_1 \left( E_{+z} \left( \hat{L}_2 E_{-z} \right)^* + E_{-z}^* \hat{L}_2 E_{+z} \right)
\end{alignat*}
Evaluating at \( z < 0 \) gives:
\begin{alignat*}{2}
	\frac{1}{v_g} &\frac{\partial E}{\partial t}(z,t) + \frac{\partial E}{\partial z}(z,t) = \frac{1}{v_g} \frac{\partial E_-}{\partial t}(-z,t) - \frac{\partial E_-}{\partial z}(-z,t) \\
	&= \frac{g_0}{2} \hat{L}_2 \frac{1}{\sqrt{D_0^2 - |D_2|^2}} \bigg( E_-(-z,t) + E_+(-z,t) \frac{1}{D_2^{- *}} \left( \scalebox{1}{$\sqrt{D_0^2 - |D_2|^2}$} - D_0 \right) \bigg) \\
	&= \frac{g_0}{2} \hat{L}_2 \frac{1}{\sqrt{D_0^2 - |D_2|^2}} \left( E_{+z} + E_{-z} \frac{1}{D_2^{- *}} \left( \scalebox{1}{$\sqrt{D_0^2 - |D_2|^2}$} - D_0 \right) \right)
\end{alignat*}
\begin{alignat*}{2}
	D_0 &= 1 + \frac{1}{P_s} \operatorname{Re} \hat{L}_1 \left( E^*_+(-z,t) \hat{L}_2 E_+(-z,t) + E_-^*(-z,t) \hat{L}_2 E_-(-z,t) \right) \\
	&= 1 + \frac{1}{P_s} \operatorname{Re} \hat{L}_1 \left( E^*_{-z} \hat{L}_2 E_{-z} + E^*_{+z} \hat{L}_2 E_{+z} \right) \\
	D_2^{-} &= \frac{1}{P_s} \hat{L}_1 \left( E_-(-z,t) \left(\hat{L}_2 E_+(-z,t) \right)^* + E_+^*(-z,t) \hat{L}_2 E_-(-z,t) \right) \\
	&= \frac{1}{P_s} \hat{L}_1 \left( E_{+z} \left(\hat{L}_2 E_{-z} \right)^* + E_{-z}^* \hat{L}_2 E_{+z} \right)
\end{alignat*}
Both results are identical, so we can use the lighter notation
\begin{alignat*}{2}
	E_{+z} &= E(z,t) \to E \\
	E_{-z} &= E(-z,t) \to E_-
\end{alignat*}
to combine them into a unified master equation
\begin{align}
	\frac{1}{v_g} \frac{\partial E}{\partial t} + \frac{\partial E}{\partial z} &= \frac{g_0}{2} \hat{L}_2 \frac{1}{\sqrt{D_0^2 - |D_2|^2}} \left( E + E_- \frac{1}{D_2^{*}} \left( \scalebox{1}{$\sqrt{D_0^2 - |D_2|^2}$} - D_0 \right) \right). \\
	D_0 &= 1 + \frac{1}{P_s} \operatorname{Re} \hat{L}_1 \left( E^* \hat{L}_2 E + E^*_- \hat{L}_2 E_- \right) \nonumber \\
	D_2 &= \frac{1}{P_s} \hat{L}_1 \left( E (\hat{L}_2 E_-)^* + E^*_- \hat{L}_2 E \right) \nonumber 
\end{align}

Next, we perform the CW normalization, defining the \textbf{normalized field} $F(z,t)$ in terms of the extended field $E$ and the CW field squared profile $P(z)$ (which is proportional to intensity). To compute $P(z)$, we note that the nonlinear gain in the absence of time-variation can be found by using \( \hat{L}_1 = \hat{L}_2 = \hat{I} \), giving 
\begin{align*}
	D_0 &= 1 + \frac{1}{P_s} \left( |E_+|^2 + |E_-|^2 \right) \equiv 1 + \frac{1}{P_s} (P_+ + P_-) \\
	D_2 &= \frac{2}{P_s} \left( E_\pm E_\mp^* \right) \rightarrow |D_2|^2 =  \tfrac{4}{P_s^2} \left( P_+ P_- \right) \\
	\nlgainvar_\text{NL}^\pm &= \frac{g_0}{2} \frac{P_s}{2 P_\pm} \left( 1 + \frac{P_\pm - P_\mp - P_s}{\sqrt{(P_\pm + P_\mp + P_s)^2 - 4 P_\pm P_\mp}} \right) E_\pm \\
	&= \frac{g_0}{2} \left( 1 - \frac{P_\pm + 2 P_\mp}{P_s} + \dots \right) E_\pm.
\end{align*}
where we have introduced $P_\pm \equiv |E_\pm|^2$, and $P$ is the extended version found from combining the two. Thus, we can find the steady-state CW field squared as the solution to
\begin{align}
	\pm \frac{\partial P_\pm}{\partial z} = -\alpha(z) P_\pm + g_0 \frac{P_s}{2 P_\pm} \scalebox{1.1}{$\left( 1 + \frac{P_\pm - P_\mp - P_s}{\sqrt{(P_\pm + P_\mp + P_s)^2 - 4 P_\pm P_\mp}} \right)$} P_\pm \label{cwpower}
\end{align}
where \( \alpha(z) \) is used to capture any spatially-varying losses, including mirror losses. This is an ODE and can therefore be solved rapidly. Approximate analytical solutions are also well-known. This can be expressed in the extended space as
\[
\frac{\partial P}{\partial z} = -\alpha(z) P + \underbrace{g_0 \frac{P_s}{2 P} \left( 1 + \frac{P - P_- - P_s}{\sqrt{(P + P_- + P_s)^2 - 4 P P_-}} \right)}_{\equiv g_{cw}(z)} P,
\]
where we have introduced $g_{cw}(z)$ as the effective spatially-dependent gain for CW light. Defining a dimensionless power as \( K \equiv P / P_0 \) (where \( P_0 \equiv P(z \to 0^+) \)) and \( K_s \equiv P_s / P_0 \), we get:
\[
g_{cw}(z) = g_0 \frac{K_s}{2 K} \left( 1 + \frac{K - K - K_s}{\sqrt{(K + K_- + K_s)^2 - 4 K K_-}} \right)
\]
and the evolution of \( K \) becomes:
\[
\frac{\partial K}{\partial z} = -\alpha(z) K + g_{cw}(z) K
\]
Defining \( E \equiv \sqrt{K} F \), we get:
\begin{align*}
	\frac{\partial E}{\partial z} &= \frac{1}{2} K^{-1/2} \frac{\partial K}{\partial z} F + \sqrt{K} \frac{\partial F}{\partial z} = \sqrt{K} \left( \frac{\partial F}{\partial z} + \frac{1}{2 K} \frac{\partial K}{\partial z} F \right) \\
	\frac{\partial E}{\partial t} &= \sqrt{K} \frac{\partial F}{\partial t} \implies \hat{L}_i E = \sqrt{K} \left( \hat{L}_i F \right)
\end{align*}
We investigate the normalized fields in the exact and low-field limit. The extended master equation becomes
\begin{alignat*}{2}
	\frac{1}{v_g} \frac{\partial E}{\partial t} + \frac{\partial E}{\partial z} &= -\frac{\alpha}{2} E + \frac{g_0 }{2} \hat{L}_2 \frac{1}{\sqrt{D_0^2 - |D_2|^2}} \left( E + E_- \frac{1}{D_2^*} \left( \sqrt{D_0^2 - |D_2|^2} - D_0 \right) \right) \\
	&= \frac{1}{v_g} \sqrt{K} \frac{\partial F}{\partial t} + \sqrt{K} \left( \frac{\partial F}{\partial z} + \frac{1}{2K} \frac{\partial K}{\partial z} F \right)
\end{alignat*}
with
\begin{alignat*}{2}
	g_{cw}(z) &= g_0 \frac{K_s}{2 K} \left( 1 + \frac{K - K_- - K_s }{\sqrt{(K + K_- + K_s)^2 - 4 K K_-}} \right) \\
	\frac{\partial K}{\partial z} &= - \alpha K + g_{cw}(z) K
\end{alignat*}
Dividing out \( \sqrt{K} \), we get:
\begin{alignat*}{2}
	\frac{1}{v_g} &\frac{\partial F}{\partial t} + \frac{\partial F}{\partial z} + \frac{1}{2K} \frac{\partial K}{\partial z} F = \frac{1}{v_g} \frac{\partial F}{\partial t} + \frac{\partial F}{\partial z} + \frac{1}{2K} \left( -\alpha + g \right) K F \\
	&= -\frac{\alpha}{2}F + \frac{g_0 }{2} \hat{L}_2 \frac{1}{\sqrt{D_0^2 - |D_2|^2}} \left( F + \frac{F_- \sqrt{K_-}}{\sqrt{K}} \frac{1}{D_2^*} \left( \sqrt{D_0^2 - |D_2|^2} - D_0 \right) \right)
\end{alignat*}
Introducing the intensity-like operators $S_{ij} \equiv \frac{1}{2} \hat{L}_1 \left( F_i^* \hat{L}_2 F_j + F_j \left( \hat{L}_2 F_i \right)^* \right)$,
\begin{alignat*}{2}
	S_{++} &= \frac{1}{2} \hat{L}_1 \left( F^* \hat{L}_2 F + F \left( \hat{L}_2 F \right)^* \right) = \text{Re} \, \hat{L}_1 \left( F^* \hat{L}_2 F \right) \\
	S_{--} &= \frac{1}{2} \hat{L}_1 \left( F_-^* \hat{L}_2 F_- + F_- \left( \hat{L}_2 F_- \right)^* \right) = \text{Re} \, \hat{L}_1 \left( F_-^* \hat{L}_2 F_- \right) \\
	S_{-+} &= \frac{1}{2} \hat{L}_1 \left( F_-^* \hat{L}_2 F + F \left( \hat{L}_2 F_- \right)^* \right)
\end{alignat*}
we have
\begin{alignat*}{2}
	D_0 &= 1 + \frac{1}{P_s} \operatorname{Re} \hat{L}_1 \left( \sqrt{K} F^* \hat{L}_2 \sqrt{K} F + \sqrt{K_-} F_-^* \hat{L}_2 \sqrt{K} F_- \right) \\
	&= 1 + \frac{1}{P_s} \operatorname{Re} \hat{L}_1 \left( K F^* \hat{L}_2 F + K_- F_-^* \hat{L}_2 F_- \right) \\
	&= 1 + \frac{1}{P_s} (K S_{++} + K_- S_{--}) \\
	D_2 &= \frac{1}{P_s} \hat{L}_1 \left( \sqrt{K} F \left( \hat{L}_2 \sqrt{K_-} F_- \right)^* + \sqrt{K_-} F_-^* \left( \hat{L}_2 \sqrt{K} F \right) \right) \\
	&= \frac{1}{P_s} \sqrt{K K_-} \hat{L}_1 \left( F( \hat{L}_2 F_-)^* + F_-^* ( \hat{L}_2 F ) \right) \\
	&= \frac{2}{P_s} \sqrt{K K_-} S_{-+}
\end{alignat*}
and
\begin{alignat*}{2}
	\frac{1}{v_g} \frac{\partial F}{\partial t} + \frac{\partial F}{\partial z} &= \frac{g_0}{2} \hat{L}_2 \frac{1}{\sqrt{D_0^2 - |D_2|^2}} \left( F + \frac{F_- \sqrt{K_-}}{\sqrt{K}} \frac{1}{D_2^*} \left( \sqrt{D_0^2 - |D_2|^2} - D_0 \right) \right) \\
	&\quad - \frac{g_{cw}(z)}{2} F
\end{alignat*}
Using $\frac{F_- \sqrt{K_-}}{\sqrt{K}} \frac{1}{\frac{2}{P_s} \sqrt{K K_-} S_{-+}^*} = \frac{P_s}{2K} \frac{F_-}{S_{-+}^*}$,
\begin{align}
	&\frac{1}{v_g} \frac{\partial F}{\partial t} + \frac{\partial F}{\partial z} = \frac{g_0}{2} \hat{L}_2 \frac{1}{\scalebox{0.8}{$\sqrt{D_0^2 - |D_2|^2}$}} \left( F + \frac{P_s}{2K} \frac{F_-}{S_{-+}^*} \left( \scalebox{0.8}{$\sqrt{D_0^2 - |D_2|^2}  - D_0$} \right) \right)  - \frac{g_{cw}(z)}{2} F \label{exactnorm0} \\
	&\quad\quad  D_0 = 1 + \frac{1}{P_s} \left( K S_{++} + K_- S_{--} \right) \nonumber \\
	&\quad\quad  D_2 = \frac{2}{P_s} \sqrt{K K_-} S_{-+} \nonumber  \\
	&\quad\quad  S_{ij} \equiv \frac{1}{2} \hat{L}_1 \left( F_i^* \hat{L}_2 F_j + F_j \left( \hat{L}_2 F_i \right)^* \right) \nonumber 
\end{align}
This can be rewritten more compactly by introducing the dimensionless field $f \equiv F/\sqrt{P_0}$, normalized CW field squared $k \equiv K / K_s$, and intensity functions as $s_{ij} \equiv \frac{1}{2} \hat{L}_1 ( f_i^* \hat{L}_2 f_j + f_j ( \hat{L}_2 f_i )^* )$. In this case, $E(z,t)=E_s \sqrt{k(z)}f(z,t)$, and we at last have an exact normalized master equation describing bidirectional fields:
\begin{empheq}[box=\tcbhighmath]{align*}
	&\frac{1}{v_g} \frac{\partial f}{\partial t} + \frac{\partial f}{\partial z} = \frac{g_0}{2} \hat{L}_2 \frac{1}{\scalebox{0.8}{$\sqrt{D_0^2 - |D_2|^2}$}} \left( f + \frac{1}{2k} \frac{f_-}{s_{-+}^*} \left( \scalebox{0.8}{$\sqrt{D_0^2 - |D_2|^2}  - D_0$} \right) \right)  - \frac{g_{cw}(z)}{2} f \\
	&\quad\quad  D_0 = 1 + k s_{++} + k_- s_{--}  \\
	&\quad\quad  D_2 = 2 \sqrt{k k_-} s_{-+} \\
	&\quad\quad  s_{ij} \equiv \frac{1}{2} \hat{L}_1 \left( f_i^* \hat{L}_2 f_j + f_j \left( \hat{L}_2 f_i \right)^* \right) \\
	&\text{where $k(z)$ is the solution to} \\
	&\quad\quad  \frac{\partial k}{\partial z} = (-\alpha(z) + g_{cw}(z)) k \\
	&\quad\quad  g_{cw}(z) = g_0 \frac{1}{2k} \left( 1 + \frac{k - k_- - 1}{\scalebox{1}{$\sqrt{(1 + k + k_- )^2 - 4 k k_-}$}} \right) 
\end{empheq}
\vspace{-10mm}
\begin{align} \label{exactnorm}
\end{align}
In the case of a unidirectional wave, $f_-=0$, and the normalized master equation reduces to 
\begin{align}
	\frac{1}{v_g} \frac{\partial f}{\partial t} + &\frac{\partial f}{\partial z} = \frac{g_0}{2} \hat{L}_2 \frac{1}{1 + k \operatorname{Re} \hat{L}_1 ( f^* \hat{L}_2 f )  } f - \frac{g_{cw}(z)}{2} f \nonumber \\
	&g_{cw}(z) = g_0 \frac{1}{1+k}
\end{align}
This agrees with the familiar result for unidirectional lasers, but importantly is general for the non-adiabatic case and properly orders the $\hat{L}_1$ and $\hat{L}_2$ operators.

When (\ref{exactnorm0}) is expanded in the low-field ($\mathcal{O}(E^3)$) and adiabatic ($\mathcal{O}(\frac{\partial}{\partial t})$) limit, one recovers
\begin{align*}
	\frac{1}{v_g} \frac{\partial F}{\partial t} &+ \frac{\partial F}{\partial z} = -\frac{g_0}{2 P_s} \Bigg[ K \Big( |F|^2 - P_0 - (T_1 + \tfrac{3}{2} T_2) \frac{\partial F^*}{\partial t}F - (T_1 + \tfrac{5}{2} T_2) |F|^2 \frac{\partial}{\partial t} \Big) \\
	&\quad + K_- \Big(2 (|F_-|^2 - P_0) - (T_1 + \tfrac{5}{2} T_2) \scalebox{1}{$ \left( \frac{\partial F_-^*}{\partial t} F_- + F_-^* \frac{\partial F_-}{\partial t} \right)$} - 2 T_2 |F_-|^2 \frac{\partial}{\partial t} \Big) \\
	&\quad - K_- \Big( T_1 + \frac{1}{2} T_2 \Big) \scalebox{1}{$\Big( \frac{\partial F^*}{\partial t} F_- + |F_-|^2 \frac{\partial}{\partial t} \Big)$} \Bigg] F \\
	&\quad + \frac{g_0}{2} (\hat{L}_2 - 1) F
\end{align*}
where the first line is solely from self-saturation, the second is from cross-saturation, the third from cross-steepening, and the last from the lineshape. This matches prior adiabatic results.

While (\ref{exactnorm}) is exact, it contains a nonlinear dependence on counterpropagating fields that hinder the construction of a mean-field theory. However, a nearly-exact non-adiabatic mean-field theory can still be constructed if \( k \) is expanded to first-order (equivalent to expanding E to third-order), resulting in
\begin{alignat*}{2}
	\frac{1}{v_g} \frac{\partial f}{\partial t} + \frac{\partial f}{\partial z} &\approx \frac{g_0}{2} \hat{L}_2 \left[ f -   \left(k s_{++} + k_- s_{--}\right) f - k_- s_{-+} f_-  \right] - \frac{g_0}{2} \left( 1 - (k+2 k_-) \right) f
\end{alignat*}
Importantly, this expansion can be used in conjunction with the \textit{exact} value of $k$ (determined by (\ref{cwpower})) to capture the important dynamics of the system. Note that for the CW solution, which has $f=1$ and all $s_{ij}=1$, the right hand side vanishes and one recovers an undisturbed traveling wave. Rewriting this, 
\begin{align*}
	\frac{1}{v_g} \frac{\partial f}{\partial t} + \frac{\partial f}{\partial z} = -&\frac{g_0}{2} \hat{L}_2 \left[ (k s_{++} + k_- s_{--}) f + k_- s_{-+} f_- \right] \\
	+ &\frac{g_0}{2} \left(\hat{L}_2 - 1 + k + 2k_- \right) f
\end{align*}
where $k$ is computed from  (\ref{cwpower}). The first two terms on the first line are respectively the result of self-saturation and cross-saturation, while the third is due solely to backscattering. In steady state, the backscattering term produces a second cross-saturation factor, which is ultimately responsible for the factor of 2 on the $|F_-|^2$ term.

Lastly, we note that this formalism contains no explicit dependence on $T_1$ or $T_2$ except for that dependence contained within $\hat{L}_1$ and $\hat{L}_2$. While both the saturation power $P_s$ and small-signal gain $g_0$ do contain these factors, they are not relevant for the normalized theory. The saturation power has actually dropped out of the normalized theory completely, only affecting the final power scale value. The dependence inside $g_0$ is generally ignored, as a laser system's gain must overcome its losses, and so it is more prudent to treat the small-signal gain as an independent pump variable.

\section{Mean-field averaging} \label{mfaverage}
For a nonlinear gain of the form
\begin{align*} 
	\frac{1}{v_g} \frac{\partial f}{\partial t} + \frac{\partial f}{\partial z} &= -\frac{g_0}{2} \hat{L}_2 \Big[ (k s_{++} + k_- s_{--}) f + k_- s_{-+} f_- \Big] \notag\\
	&\quad + \frac{g_0}{2} \left( \hat{L}_2 - 1 + (k + 2 k_-) \right) f \\
	&= \frac{1}{2} \hat{g}_{\textrm{NL}} f = \nlgainvar(f, f_-)
\end{align*} 
we would like to find the mean gain
\begin{align*} 
	\langle \nlgainvar \rangle = \frac{1}{T_r} \int_0^{T_r} \nlgainvar (z + v_g t, z, t) \, dt
\end{align*} 
where \( T_r=L_r / v_g \) is the cavity round-trip time and \( L_r \) is the round-trip length. In the extended cavity we expect waves that are primarily positive-traveling and take the form
\begin{align*} 
	f(z, t) &= f(z - v_g t) \\
	f_-(z, t) &= f(-z, t) = f(-z - v_g t)
\end{align*} 
so the mean can be found as
\begin{align*} 
	\langle \nlgainvar \rangle = \frac{1}{T_r} \int_0^{T_r} \big[  \nlgainvar \left( f(z - v_g t), f(-z - v_g t) \right) \big]_{z \to z + v_g t \atop t \to t} \, dt
\end{align*} 
Substituting \( u \equiv -(z - v_g t) \), \( du = v_g \, dt \), and \( t = \frac{u + z}{-v_g} \), and exploiting periodicity to change the limits of integration to
\begin{align*} 
	\int_0^{T_r} dt = \frac{1}{v_g} \int_{-z}^{-(z - v_g T_r)} du = \frac{1}{v_g} \int_0^{L_r} du
\end{align*} 
leads to
\begin{align*} 
	\langle \nlgainvar \rangle = \frac{1}{L_r} \int_0^{L_r} \nlgainvar \big[ f(z - v_g t), f(-z - v_g t) \big]_{z \to -u \atop t \to -\frac{1}{v_g} (u + z)} \, du.
\end{align*} 
or more explicitly,
\begin{align*} 
	\nlgainvar_w (z,t) &\equiv \nlgainvar \left( f(z - v_g t), f(-z - v_g t) \right) \\
	\langle \nlgainvar \rangle& = \frac{1}{L_r} \int_0^{L_r} \nlgainvar_w \left( -u, -\frac{1}{v_g} (u + z) \right) \, du
\end{align*} 
Note that the substitution rules can be more conveniently expressed as
\begin{align*} 
	z - v_g t &\to -u - v_g \left( -\tfrac{1}{v_g} \right) (u + z) = z \\
	-z - v_g t &\to u - v_g \left( -\tfrac{1}{v_g} \right) (u + z) = z + 2 u
\end{align*} 
Some example mean-field averages are computed below:
\begin{enumerate}
	\item Linear gain of the form \(\nlgainvar (z, t) = \frac{g_0}{2} (\hat{L}_2 - 1) f\):
	\begin{align*} 
		\nlgainvar_w (z, t) &= \frac{g_0}{2} (\hat{L}_2 - 1) f(z - v_g t) = \frac{g_0}{2} \left( -T_2 \frac{\partial}{\partial t} + \dots \right) f(z - v_g t) \\
		&= \frac{g_0}{2} \left( v_g T_2 \frac{\partial}{\partial z} + \dots \right) f(z - v_g t) = \frac{g_0}{2} (\hat{\mathcal{L}}_{2} - 1) f(z - v_g t) \\
		\langle \nlgainvar \rangle &= \frac{1}{L_r} \int_0^{L_r} d u \, \frac{g_0}{2} \left( (\hat{\mathcal{L}}_{2} - 1) f \right) (z) = \frac{g_0}{2} \left( \hat{\mathcal{L}}_{2} - 1 \right) f
	\end{align*} 
	where we have introduced \(\hat{\mathcal{L}}_{2}\equiv(1-v_g T_2 \partial z)^{-1}\) as the position-space version of \(\hat{L}_2\). Since both \( f(z - v_g t) \) and \( f(-z - v_g t) \) have the same time dependence, \( \partial_t \) acting on one of them will always convert to \( -v_g \partial_z \).
	\item Cross-saturation of the form \(\nlgainvar (z, t) = -\frac{g_0}{2} k_- |f_-|^2 f \):
	\begin{align*} 
		\nlgainvar_w (z, t) &= -\frac{g_0}{2} k(-z) |f(-z - v_g t)|^2 f(z - v_g t) \\
		\langle \nlgainvar \rangle &= \frac{1}{L_r} \int_0^{L_r}  -\frac{g_0}{2} k(u) |f(z + 2u)|^2 f(z)  du = -\frac{g_0}{2} \tilde{k} \left[ |f|^2 \right] f
	\end{align*} 
	where we have introduced the convolution with \( k \) as
	\begin{align*} 
		\tilde{k} \left[ f \right] (z) &\equiv \frac{1}{L_r} \int_0^{L_r} k(u) f(z + 2u) du .
	\end{align*} 
	It is also useful to define the mean value of \( k \) as \(\langle k \rangle \equiv \frac{1}{L_r} \int_0^{L_r} k(u) du\), since when positive and negative traveling waves separate, general terms involving these waves will transform like
	\begin{align*} 
		k_- f_- g_+ &\to \tilde{k} \left[ f \right] g \\
		k g_+ &\to \langle k \rangle g
	\end{align*} 
\end{enumerate}
The main difficulty with the exact master equation formulation is that \( f \) and \( f_- \) do not usually separate. For example, standard gain saturation would require the computation of
\begin{align*} 
	\frac{1}{L_r} \int_0^{L_r} du \, \left[ 1 + |f(z)|^2 + |f(z + 2u)|^2 \right]^{-1}
\end{align*} 
which is not easily calculated. Therefore, series expansions such as the product expansion of Lorentzians are still needed to separate counterpropagating terms. We will split the low-field expression into saturation and backscattering components, denoted by \(\nlgainvar_s\) and \(\nlgainvar_x\) as:
\begin{align*} 
	\nlgainvar &= \nlgainvar_s + \nlgainvar_x \\
	\nlgainvar_s &\equiv -\frac{g_0}{2} \hat{L}_2 \left[ \left( k s_{++} + k_- s_{--} \right) f \right] + \frac{g_0}{2} \left( \hat{L}_2 - 1 + k + 2k_- \right) f \\
	\nlgainvar_x &\equiv -\frac{g_0}{2} \hat{L}_2 k_- s_{-+} f_-
\end{align*} 

\subsection{Backscattering term}
The backscattering term is FM-generating but is also the most difficult to calculate, as it is ultimately created by the interaction with counterpropagating waves
\begin{align*} 
	\nlgainvar_x &= -\frac{g_0}{2} \hat{L}_2 k_- f_- s_{-+} \\
	&= -\frac{g_0}{4} \hat{L}_2 k_- f_- \hat{L}_1 \left( f_-^* \hat{L}_2 f + f( \hat{L}_2 f_- )^* \right)
\end{align*} 
Each term is of the form \(\hat{L}_2 f \hat{L}_1 (gh)\), where \(f\) and \(g\) travel in the negative direction and \(h\) in the positive direction. We will use the Lorentzian product identity (equation (\ref{productexpansion}))
\begin{align*} 
	\hat{L_i}(fg) = \sum_{n=0}^\infty \left(\hat{B_i}_n f \right) \left(\hat{B_i}_n g \right) \quad \text{where} \quad 
	\hat{B_i}_n=\frac{\left( T_i \frac{\partial}{\partial t} \right)^n}{\left( 1 + T_i \frac{\partial}{\partial t} \right)^{n+1}}
\end{align*} 
to combine the two negative traveling wave terms:
\begin{align*} 
	\hat{L}_2 f \hat{L}_1 (gh) &= \hat{L}_2 f \sum_n \left( \hat{B}_{1n} g \right) \left( \hat{B}_{1n} h \right) \\
	&= \sum_n \hat{L}_2 f \left( \hat{B}_{1n} g \right) \left( \hat{B}_{1n} h \right) \\
	&= \sum_{n,m} \left( \hat{B}_{2m} f \, \hat{B}_{1n} g \, \right) \left( \hat{B}_{2m} \hat{B}_{1n} h  \right)
\end{align*} 
\begin{align*} 
	\nlgainvar_x = - \frac{g_0}{4} k_- \sum_{n,m} &\left( \hat{B}_{2m} f_- \hat{B}_{1n} f_-^* \right) \left( \hat{B}_{2m} \hat{B}_{1n} \hat{L}_2 f \right) \\
	\quad + &\left( \hat{B}_{2m} f_- \hat{B}_{1n} ( \hat{L}_2 f_- )^* \right) \left( \hat{B}_{2m} \hat{B}_{1n} \hat{L}_2 f \right)
\end{align*} 
Since all of the operators now apply exclusively to like-direction waves, we can now perform a mean-field average in the usual way (applying \(\tilde{k}\) to negative traveling waves and converting \(\partial_t \rightarrow -v_g \partial_z\)). Defining \(\hat{\mathcal{B}}_{ik} \equiv \hat{B}_{ik}(\partial_t \rightarrow -v_g \partial_z) \), the mean-field expression becomes
\begin{align*} 
	\langle \nlgainvar_x \rangle = - \frac{g_0}{4} \sum_{n,m=0}^{\infty} &\tilde{k} \left[ \hat{\mathcal{B}}_{2m} f \hat{\mathcal{B}}_{1n} f^* \right] \hat{\mathcal{B}}_{2m} \hat{\mathcal{B}}_{1n} \hat{\mathcal{L}}_2 f \\
	\quad + &\tilde{k} \left[ \hat{\mathcal{B}}_{2m} f \hat{\mathcal{B}}_{1n} \hat{\mathcal{L}}_2 f^* \right] \hat{\mathcal{B}}_{2m} \hat{\mathcal{B}}_{1n} f 
\end{align*} 

\subsection{Gain saturation term}
The saturation term is easier to deal with than backscattering, but is not completely trivial. It has
\begin{align*} 
	\nlgainvar_s &= -\frac{g_0}{2} \hat{L}_2 \left[ (k s_{++} + k_- s_{--}) f \right] + \frac{g_0}{2} \left( \hat{L}_2 - 1 + k + 2k_- \right) f
\end{align*} 
where
\begin{align*} 
	s_{++} &= \operatorname{Re} \hat{L}_1 \left( f^* \hat{L}_2 f \right) \\
	s_{--} &= \operatorname{Re}  \hat{L}_1 \left( f_-^* \hat{L}_2 f_- \right)
\end{align*} 
Letting \( s \equiv  \operatorname{Re}  \hat{\mathcal{L}}_1 f^* \hat{\mathcal{L}}_2 f \) (which is intensity-like, as \( s \sim |f|^2 \)):
\begin{align*} 
	\nlgainvar_{sw} (z,t) = \frac{g_0}{2} \hat{L}_2 \Big( &k(z) \operatorname{Re} \hat{L}_1 \left( f^*(z - v_g t) \hat{L}_2 f(z - v_g t) \right) \\
	\quad + &k(-z) \text{Re} \hat{L}_1 \left( f^*(-z - v_g t) \hat{L}_2 f(-z - v_g t) \right) \Big) \\
	&+ \frac{g_0}{2} \left( \hat{L}_2 - 1 + (k + 2k_-) \right) f(z - v_g t)
\end{align*} 
and
\begin{align*} 
	\hat{L}_1 f^*(z - v_g t) \hat{L}_2 f(z - v_g t) &= \hat{L}_1 f^*(z - v_g t)(\hat{\mathcal{L}}_2 f)(z - v_g t) \\
	&= s(z - v_g t) \\
	\hat{L}_1 f^*(-z - v_g t) \hat{L}_2 f(-z - v_g t) &= \hat{L}_1 f^*(-z - v_g t)(\hat{\mathcal{L}}_2 f)(-z - v_g t) \\ 
	&= s(-z - v_g t) \equiv s_-.
\end{align*} 
The time-domain expression becomes
\begin{align*} 
	\nlgainvar_{sw}(z,t) &= -\frac{g_0}{2} \hat{L}_2 \left( k s f + k_- s_- f \right) + \frac{g_0}{2} \left( \hat{L}_2 - 1 + (k + 2k_-) \right) f \\
	&= -\frac{g_0}{2}  \left( k \hat{L}_2 s f + k_- \sum_m ( \hat{B}_{2m} s_- ) ( \hat{B}_{2m} f ) \right) \\
	&\quad+ \frac{g_0}{2} \left( \hat{L}_2 - 1 + k + 2k_- \right) f
\end{align*} 
Taking the mean-field average yields
\begin{align*} 
	\langle \nlgainvar_s \rangle &= -\frac{g_0}{2} \left( \langle k \rangle \hat{\mathcal{L}}_2 (s f) + \sum_{m=0}^{\infty} \tilde{k} \left[ \hat{\mathcal{B}}_{2m} s \right] \hat{\mathcal{B}}_{2m} f \right) \\
	&\quad+ \frac{g_0}{2} \left( (\hat{\mathcal{L}}_2 - 1) f + 3 \langle k \rangle \right) \\
	\langle \nlgainvar_s \rangle&= -\frac{g_0}{2} \left( \langle k \rangle \hat{\mathcal{L}}_2 s + \sum_{m=0}^{\infty} \tilde{k} \left[ \hat{\mathcal{B}}_{2m} s \right] \hat{\mathcal{B}}_{2m} - 3 \langle k \rangle  \right) f 
	+ \frac{g_0}{2} (\hat{\mathcal{L}}_2 - \hat{I}) f
\end{align*} 

\subsection{Full operator}

Putting it all together leads to the following nonlinear mean-field gain operator for bidirectional cavities:
\begin{empheq}[box=\tcbhighmath]{align*}
	\frac{1}{v_g} \frac{\partial f}{\partial T} &= \frac{1}{2} \langle \hat{g}_{\textrm{NL}} \rangle f \\
	\frac{\langle \hat{g}_{\textrm{NL}} \rangle}{g_0} &= \hat{\mathcal{L}}_2 - \hat{I} - \bigg( \langle k \rangle \hat{\mathcal{L}}_2 s + \sum_{m=0}^{\infty} \tilde{k} \left[ \hat{\mathcal{B}}_{2m} s \right] \hat{\mathcal{B}}_{2m} - 3 \langle k \rangle  \\
	&\hspace{2cm} + \frac{1}{2} \sum_{n,m=0}^{\infty} \tilde{k} \left[ \hat{\mathcal{B}}_{2m} f \, \hat{\mathcal{B}}_{1n} f^* \right] \hat{\mathcal{B}}_{2m} \hat{\mathcal{B}}_{1n} \hat{\mathcal{L}}_2 +
	\tilde{k} \left[ \hat{\mathcal{B}}_{2m} f \, \hat{\mathcal{B}}_{1n} \hat{\mathcal{L}}_2 f^* \right] \hat{\mathcal{B}}_{2m} \hat{\mathcal{B}}_{1n} \bigg) \\
	\hat{\mathcal{L}}_i &\equiv \frac{1}{1 - v_g T_i \frac{\partial}{\partial z}} \\
	\hat{\mathcal{B}}_{ik} &\equiv  \frac{\left( -v_g T_i \frac{\partial}{\partial z} \right)^k}{\, \left(1 - v_g T_i \frac{\partial}{\partial z} \right)^{k+1}} = (\hat{\mathcal{L}}^{-1}_i-1)^k \hat{\mathcal{L}}^{k+1}_i \\
	s &\equiv \operatorname{Re} \hat{\mathcal{L}}_1 \left( f^* \hat{\mathcal{L}}_2 f \right)
\end{empheq} 
\vspace{-5mm}
\begin{align} \label{mfresult}
\end{align}
The first line in this expansion leads to amplitude relaxation (driving the amplitude back to unity), while the second is responsible for backscattering. The series are unconditionally convergent, and each operator is well-behaved numerically (automatically low-pass filtering high frequencies). While the product series are included for exactness, in most cases only the first term is necessary and one can write
\begin{align*}
	\frac{\hat{g}_{\textrm{NL}}}{g_0} &= \hat{\mathcal{L}}_2 - \hat{I} - \bigg( \langle k \rangle \hat{\mathcal{L}}_2 s + \tilde{k} \left[ \hat{\mathcal{L}}_{2} s \right] \hat{\mathcal{L}}_{2} - 3 \langle k \rangle  \\
	&\hspace{2cm} + \frac{1}{2} \big( \tilde{k} \left[ \hat{\mathcal{L}}_{2} f \, \hat{\mathcal{L}}_{1} f^* \right] \hat{\mathcal{L}}_{2} \hat{\mathcal{L}}_{1} \hat{\mathcal{L}}_2 
	+ \tilde{k} \left[ \hat{\mathcal{L}}_{2} f \, \hat{\mathcal{L}}_{1} \hat{\mathcal{L}}_2 f^* \right] \hat{\mathcal{L}}_{2} \hat{\mathcal{L}}_{1} \bigg)
\end{align*} 
In the adiabatic limit, this reduces to 
\begin{align*} 
	\frac{\hat{g}_{\textrm{NL}}}{g_0} &=  \hat{\mathcal{L}}_2 - \hat{I} - \langle k \rangle |f|^2 - 2 \tilde{k} \left[ |f|^2 \right] + 3 \langle k \rangle  \\
	&\quad + v_g \left( 2 T_1 + 3 T_2 \right) \tilde{k} \left( f \frac{\partial f^*}{\partial z} \right) + v_g \left( T_1 + \frac{3}{2} T_2 \right) \langle k \rangle \frac{\partial f^*}{\partial z} f \\
	&\quad + v_g \left( T_1 + \frac{5}{2} T_2 \right) \left( \tilde{k} \left[ |f|^2 \right] \frac{\partial}{\partial z} + \langle k \rangle |f|^2 \frac{\partial}{\partial z} + \tilde{k} \left[ f^* \frac{\partial f}{\partial z} \right] \right)
\end{align*} 
which agrees with the prior result. 

For the case of a unidirectional wave, one can simply remove the backscattering interactions, and one arrives at 
\begin{align}
	\frac{\langle \hat{g}_{\textrm{NL}} \rangle}{g_0} &= \hat{\mathcal{L}}_2 - \hat{I} - \langle k \rangle \left( \hat{\mathcal{L}}_2 s -  \hat{I}  \right).
\end{align} 
Lastly, we note that in each case, the gain expressions can instead be linearized around the CW value ($s_{ij}\approx 1$) rather than linearizing around small k. In this case, the mean-field theory is somewhat more complicated, but would possibly be more exact far above threshold.

\section{Lorentzian Factor}
\label{lorfactor}
Note that the action of the Lorentzian operator can always be written using an integrating factor as
\[
\hat{L}f = \int_{-\infty}^{t} dt' f(t') \frac{1}{T} e^{-(t - t')/T}
\]
since
\begin{alignat*}{2}
	\left( 1 + T \frac{\partial}{\partial t} \right) \hat{L}f &= \int_{-\infty}^{t} dt' f(t') \frac{1}{T} e^{-(t - t')/T} 
	+ T \bigg[ e^{-t/T} \cdot - \frac{1}{T} \int_{-\infty}^{t} dt' f(t') \frac{1}{T} e^{t'/T}  \\
	&\hspace{4.6cm}+ e^{-t/T} f(t) \frac{1}{T} e^{t/T} \bigg]\\
	&= \int_{-\infty}^{t} dt' f(t') \frac{1}{T} e^{-(t - t')/T} - \int_{-\infty}^{t} dt' f(t') \frac{1}{T} e^{-(t - t')/T} + f(t) \\
	&= f(t).
\end{alignat*}
It can also be written in terms of an impulse response  \( \ell(t) \equiv \frac{1}{T} e^{-t/T} u(t) \) as
\begin{alignat*}{2}
	\hat{L}f &= \int_{-\infty}^{\infty} dt' f(t') \frac{1}{T} e^{-(t - t')/T} \, u(-(t' - t)) \\
	&= \int_{-\infty}^{\infty} dt' f(t') \, \ell(t - t') .
\end{alignat*}
A related operator that applies to positive-traveling waves is the position-space version of $\hat{L}$, which we will denote by $\hat{\mathcal{L}}$:
\begin{alignat*}{2}
	\hat{\mathcal{L}} f &= \left( 1 - v_g T \frac{\partial}{\partial z} \right)^{-1} \\
	&= \int_{z}^{\infty} dz' f(z') \frac{1}{v_g T} e^{(z - z')/{v_g T}} \\
	\left( 1 - v_g T \frac{\partial}{\partial z} \right)\hat{\mathcal{L}} f &= \int_{z}^{\infty} dz' f(z') \frac{1}{v_g T} e^{(z - z')/{v_g T}} \\
	&- v_g T  e^{z/{v_g T}} \frac{1}{v_g T} \int_{z}^{\infty} dz' f(z') \frac{1}{v_g T} e^{-z'/{v_g T}} \\
	&+ v_g T e^{z/{v_g T}} f(z)  \frac{1}{v_g T} e^{-z/{v_g T}} = f(z).
\end{alignat*}
Thus, the operator can be written in terms of an impulse response \\
\mbox{\( \ell_z(z) = \frac{1}{v_g T} e^{z/{v_g T}} u(-z) \)} as
\[
\hat{\mathcal{L}} f = \int_{-\infty}^{\infty} dz' u(z' - z) f(z') \frac{1}{v_g T} e^{(z - z')/{v_g T}} = \int_{-\infty}^{\infty} dz' f(z') \, \ell_z(z - z') .
\]
Note that for a forward-traveling wave, the impulse response is at negative z, i.e. the effect of an impulse in field lies behind that field.

\subsection{Action on products}
To find a mean field theory, we need to be able to separate waves traveling in different directions. When \( \hat{L} \) acts on a product, the resulting term is no longer separable. However, products can computed by noting that \( \hat{L}(f g) \approx \hat{L}(f) \hat{L}(g) \) to first order:
\begin{align*}
	\hat{L}(f g) &= f g - T \frac{\partial}{\partial t}(f g) + T^2 \frac{\partial^2}{\partial t^2}(f g) + \dots \\
	&= f g - T \left( \frac{\partial f}{\partial t} g + f \frac{\partial g}{\partial t} \right) + T^2 \left( \frac{\partial^2 f}{\partial t^2} g + 2 \frac{\partial f}{\partial t} \frac{\partial g}{\partial t} + f \frac{\partial^2 g}{\partial t^2} \right) + \dots \\
	&= \sum_{n=0}^\infty (-T)^n \sum_{m=0}^n \binom{n}{m} f^{(m)} g^{(n-m)}
\end{align*}
\text{Doing a formal series:}
\begin{align*}
	\hat{L}(f g) - \hat{L}(f) \hat{L}(g) &= \sum_{n=0}^\infty (-T)^n \sum_{m=0}^n \binom{n}{m} f^{(m)} g^{(n-m)} \\
	&- \sum_{n=0}^\infty (-T)^n f^{(n)} \sum_{m=0}^\infty (-T)^m g^{(m)} \\
	&=\frac{1}{1 + T \left. \frac{\partial}{\partial t}\right|_f+ T \left. \frac{\partial}{\partial t}\right|_g}\frac{T \left. \frac{\partial}{\partial t}\right|_f}{1 + T \left. \frac{\partial}{\partial t}\right|_f} \frac{T \left. \frac{\partial}{\partial t}\right|_g}{1 + T \left. \frac{\partial}{\partial t}\right|_g}fg \\
	&=\hat{L}\left((\hat{L'}f) (\hat{L'}g)\right)
\end{align*}
where we have introduced $\hat{L'} \equiv \frac{T \partial_t }{1 + T \partial_t }$. While this form is exact, it has the same problem as the original product, and could contain mixed \( + \) and \( - \) traveling waves. However, note that \( \hat{L}' \) is essentially a high-pass filter and has reduced the spectral content of \( f \) and \( g \). The same identity can be applied iteratively to construct a series representation:
\begin{align}
	\hat{L}(f g) &= \hat{L}(f) \hat{L}(g) + \hat{L}(\hat{L}'(f)) \hat{L}(\hat{L}'(g)) + \hat{L}(\hat{L}'^{\,2}(f) \hat{L}'^{\,2}(g) ) \nonumber \\
	\hat{L}(f g) &= \sum_{n=0}^\infty \hat{L}\left(\hat{L}'^{\,n}(f)\right) \hat{L}\left(\hat{L}'^{\,n}(g)\right) \nonumber \\
	&= \sum_{n=0}^\infty \left(\hat{B}_n f \right) \left(\hat{B}_n g \right)
	\label{productexpansion}
\end{align}
where we have introduced the operators $\hat{B}_n$ as
\begin{align}
	\hat{B}_n \equiv \hat{L} \hat{L}'^{\,n} = \frac{\left( T \frac{\partial}{\partial t} \right)^n}{\left( 1 + T \frac{\partial}{\partial t} \right)^{n+1}}
\end{align}
The $\hat{B}_n$ operators can be construed as band-pass filters, as their magnitudes peak at the frequencies $\omega_n=\sqrt{n}/T$. The value of this expansion is that it separates the two waves efficiently, using well-behaved functions. It also has an \( L \) on the outside, so it works for long \( T \). It also has an infinite radius of convergence. In Fourier space, (\ref{productexpansion}) is
\begin{align*}
	&\sum_n \frac{1}{2\pi} \int F(\omega_1) \frac{1}{1 + i \omega_1 T} \left( \frac{i \omega_1 T}{1 + i \omega_1 T} \right)^n e^{i \omega_1 t} d\omega_1
	\int G(\omega_2) \frac{1}{1 + i \omega_2 T} \left( \frac{i \omega_2 T}{1 + i \omega_2 T} \right)^n e^{i \omega_2 t} d\omega_2 \\
	&= \frac{1}{(2\pi)^2} \int d\omega_1 d\omega_2 F(\omega_1) G(\omega_2)  e^{i(\omega_1 + \omega_2)t}
	\frac{1}{(1 + i \omega_1 T)(1 + i \omega_2 T)} \sum_n \left( \frac{i \omega_1 T}{1 + i \omega_1 T} \frac{i \omega_2 T}{1 + i \omega_2 T} \right)^n 
\end{align*}
Each term in the geometric series has a magnitude squared of $\frac{\omega_1 T}{1 + (\omega_1 T)^2} \frac{\omega_2 T}{1 + (\omega_2 T)^2} < 1$, so convergence is guaranteed irrespective of the size of T. Moreover, the series converges to 
\[
\frac{1}{(1 + i \omega_1 T)(1 + i \omega_2 T)} \sum_n \left( \frac{i \omega_1 T}{1 + i \omega_1 T} \frac{i \omega_2 T}{1 + i \omega_2 T} \right)^n  = \frac{1}{1 + i(\omega_1 + \omega_2) T}
\]
which agrees with a direct expansion of $\hat{L}(f g)$:
\begin{align*}
	\hat{L}(f g) &= \frac{1}{2\pi} \int \frac{d\omega}{1 + i \omega T} \left( \int dt' f(t') g(t) e^{-i \omega t'} \right) e^{i \omega t} \\
	&= \frac{1}{(2\pi)^3} \int d\omega d\omega_1 d\omega_2 dt' \frac{e^{i (\omega t - \omega_1 t' - \omega_2 t')} F(\omega_1) G(\omega_2)}{1 + i \omega T}\\
	&\hspace{2cm}\int dt' e^{i (-\omega + \omega_1 + \omega_2)t'} = 2 \pi \delta(\omega - \omega_1 - \omega_2) \\
	&= \frac{1}{(2\pi)^2} \int d\omega_1 d\omega_2 \frac{e^{i(\omega_1 + \omega_2)t} F(\omega_1) G(\omega_2)}{1 + i(\omega_1 + \omega_2) T}.
\end{align*}

\section{Modification for off-resonant injection}\label{offres}

When \( \omega_0 \) is not at the center frequency, equation (\ref{wddynamics}) becomes:
\[
\dot{d} = \frac{1}{i \hbar} \left( w \mu \right) E(t) + \left( i \omega_0 - \frac{1}{T_2} \right) d
\]
Using \( d = \tilde{d} e^{i \omega_0 t} \),
\begin{alignat*}{2}
	\dot{\tilde{d}} e^{i \omega_0 t} + i \omega_0 \tilde{d} e^{i \omega_0 t} &= \frac{1}{i \hbar} \tilde{w} \mu \left( A e^{i \omega_0 t} + A^* e^{-i \omega_0 t} \right) e^{-i \omega_0 t}+(i\omega_{eg}-\tfrac{1}{T_2})\tilde{d}e^{i \omega_0 t} \\
	\dot{\tilde{d}} &= \frac{1}{i \hbar} \tilde{w} \mu A - \left[ - i (\omega_{eg} - \omega_0) + \frac{1}{T_2} \right] \tilde{d}
\end{alignat*}
This is equivalent to letting:
\begin{alignat*}{2}
	\frac{1}{T_2} &\to \frac{1}{T_2} + i (\omega_0 - \omega_{eg}) \\
	\left( \frac{1}{T_2} + \frac{\partial}{\partial t} \right)^{-1} &\to \left( \frac{1}{T_2} + i(\omega_0 - \omega_{eg}) + \frac{\partial}{\partial t} \right)^{-1} \\
	&= T_2 \left ( 1+ T_2 \left(i(\omega_0 - \omega_{eg}) + \frac{\partial}{\partial t} \right) \right)^{-1}
\end{alignat*}
Thus, the Lorentzian operator becomes
\[
\hat{L}_2 = \left( 1 + T_2 \left( i(\omega_0 - \omega_{eg}) + \frac{\partial}{\partial t} \right) \right)^{-1}
\]
In the frequency domain, its action is
\[
\hat{L}_2(\omega) = \frac{1}{1 + i \left( \omega_0 - \omega_{eg} + \omega \right) T_2 } = \frac{1}{1 + i \left( \omega - \left(\omega_{eg} - \omega_0 \right) \right) T_2 }
\]
which behaves as expected, as it is centered at \( \omega_{eg} - \omega_0 \). However, \( \hat{L}_1 \) is unchanged.

\section{Kerr nonlinearity and dispersion}

For completeness, we include our conventions for the dispersion and Kerr nonlinearity. The Kerr nonlinearity in this formalism requires modification for the counterpropagating wave. The nonlinear polarization density can be expressed as:
\[
P_\text{NL} = \varepsilon_0 \chi^{(3)} E_f^3 = \varepsilon_0 \chi^{(3)} \left( E_+ e^{i (\omega_0 t - k_0 z)} + E_- e^{i (\omega_0 t + k_0 z)} + \text{c.c.} \right)^3
\]
The nonlinear term \( \nlgainvar_\text{NL} \) becomes:
\begin{alignat*}{2}
	\left[ \nlgainvar_\text{NL}^\pm \right]_\textrm{Kerr} &= i \frac{1}{2 \omega_0 n \varepsilon_0 c} \left( \frac{2\pi}{k_0} \right)^{-1} \int_0^{2\pi/k_0} dz \, e^{\mp i (\omega_0 t \mp k_0 z)} \frac{\partial^2 P_\text{NL}}{\partial t^2} \\
	&= - i \frac{1}{2 \omega_0 n_0 \varepsilon_0 c} \omega_0^2 \chi^{(3)} \left( 3 |E_\pm|^2 + 6 |E_\mp|^2 \right) E_\pm
\end{alignat*}
\begin{align}
\left[ \nlgainvar_\text{NL}^\pm \right]_\textrm{Kerr} &= -i \frac{3 \chi^{(3)} \omega_0}{2 n_0 c} \left( |E_\pm|^2 + 2 |E_\mp|^2 \right) E_\pm \\
	&= -i \gamma_K  \left( |E_\pm|^2 + 2 |E_\mp|^2 \right) E_\pm. \nonumber
\end{align}
Note that in our convention, the intensity of each wave is
\[
I_\pm = 2 n_0 c \varepsilon_0 |E_\pm|^2.
\]
Thus, the normalized nonlinearity can be expressed in terms of the Kerr nonlinearity as
\begin{alignat*}{2}
	\frac{3 \chi^{(3)} \omega_0}{2 n_0 c} |E_\pm|^2 &= \frac{3 \chi^{(3)} \omega_0}{2 n_0 c} \frac{I_\pm}{2 \varepsilon_0  n_0 c} = \frac{\omega_0}{c} n_2 I_\pm \, \rightarrow \, n_2 = \frac{3 \chi^{(3)} }{4 n_0^2 c \varepsilon_0} \rightarrow \gamma_K &= \frac{n_2 \omega_0}{2 n_0 \varepsilon_0}
	.
\end{alignat*}
Including second order dispersion in terms of $k''(\omega_0)$, the full master equations become
\begin{align}
	&\frac{1}{v_g} \frac{\partial E_\pm}{\partial t} \pm \frac{\partial E_\pm}{\partial z} = i \frac{1}{2} k''(\omega_0) \frac{\partial^2}{\partial t^2}E_\pm -i \gamma_K  \left( |E_\pm|^2 + 2 |E_\mp|^2 \right) E_\pm +  \nlgainvar_\text{NL}^\pm \nonumber \\
\end{align}
from which the mean-field theory is straightforwardly calculated.

\section{Glossary}

For reference, all variables and their meanings are described below:

\newcommand{\setspacing}{\setlength{\abovedisplayskip}{0pt} \setlength{\belowdisplayskip}{8pt}}
\setlength{\extrarowheight}{2pt}
\begin{longtable}{|c|p{11cm}|}
	\caption{Glossary of terms} \\
	\hline
	\textbf{Variable} & \textbf{Description} \\
	\hline
	\endfirsthead
	
	\multicolumn{2}{c}%
	{{\bfseries \tablename\ \thetable{} -- continued from previous page}} \\
	\hline
	\textbf{Variable} & \textbf{Description} \\
	\hline
	\endhead
	
	\hline \multicolumn{2}{r}{{Continued on next page}} \\
	\endfoot
	
	\hline
	\endlastfoot
	
	\multicolumn{2}{|l|}{\textbf{Fundamental Constants}} \\
	\hline
	\(\hbar\) & Reduced Planck constant \\
	\hline
	\(e\) & Elementary charge magnitude \\
	\hline
	\(\varepsilon_0\) & Vacuum permittivity \\
	\hline
	\(c\) & Speed of light in vacuum \\
	\hline
	
	\multicolumn{2}{|l|}{\textbf{Two-Level System Properties}} \\
	\hline
	\(E_g\) & Energy of the ground state \\
	\hline
	\(E_e\) & Energy of the excited state \\
	\hline
	\(\Gamma_g\) & Decay rate from the ground state \\
	\hline
	\(\Gamma_e\) & Decay rate from the excited state \\
	\hline
	\(T_1\) & Population lifetime:
	\setlength{\abovedisplayskip}{0pt}
	\setlength{\belowdisplayskip}{8pt}
	\begin{align*}
		T_1 = \dfrac{1}{\Gamma_g + \Gamma_e}
	\end{align*} \\ 
	\hline
	\(T_2\) & Coherence lifetime \\
	\hline
	\(\hat{L}_i\) & Lorentzian operator associated with lifetime \(T_i\):
	\setlength{\abovedisplayskip}{0pt}
	\setlength{\belowdisplayskip}{8pt}
	\begin{equation*}
		\hat{L}_i f = \left( 1 + T_i \dfrac{\partial}{\partial t} \right)^{-1} f = \int_{-\infty}^{t} dt' \, f(t') \dfrac{1}{T_i} e^{-(t - t')/T_i}
	\end{equation*} \\
	\hline
	\(L_i(\omega)\) & Complex Lorentzian function (lineshape function):
	\setlength{\abovedisplayskip}{0pt}
	\setlength{\belowdisplayskip}{8pt}
	\begin{equation*}
		L_i(\omega) = \dfrac{1}{1 + i \omega T_i}
	\end{equation*} \\
	\hline
	
	\multicolumn{2}{|l|}{\textbf{Density Matrix, Population, and Polarization}} \\
	\hline
	\(\rho\) & Density matrix of the two-level system \\
	\hline
	\(d\) & Coherence:
	\setlength{\abovedisplayskip}{0pt}
	\setlength{\belowdisplayskip}{8pt}
	\begin{equation*}
		d = \rho_{ge}
	\end{equation*} \\
	\hline
	\(\tilde{d}\) & Slowly varying component of the coherence:
	\setlength{\abovedisplayskip}{0pt}
	\setlength{\belowdisplayskip}{8pt}
	\begin{equation*}
		d(t) = \tilde{d}(t) e^{i \omega_0 t}
	\end{equation*} \\
	\hline
	\(w\) & Population inversion:
	\setlength{\abovedisplayskip}{0pt}
	\setlength{\belowdisplayskip}{8pt}
	\begin{equation*}
		w = \rho_{ee} - \rho_{gg}
	\end{equation*} \\
	\hline
	\(\tilde{w}\) & Slowly varying component of the population inversion:
	\setlength{\abovedisplayskip}{0pt}
	\setlength{\belowdisplayskip}{8pt}
	\begin{equation*}
		w(t) = \tilde{w}(t)
	\end{equation*} \\
	\hline
	\(w_0\) & Steady-state population inversion:
	\setlength{\abovedisplayskip}{0pt}
	\setlength{\belowdisplayskip}{8pt}
	\begin{equation*}
		w_0 = \dfrac{\Gamma_g - \Gamma_e}{\Gamma_g + \Gamma_e}
	\end{equation*} \\
	\hline
	\(\mu\) & Transition dipole moment:
	\setlength{\abovedisplayskip}{0pt}
	\setlength{\belowdisplayskip}{8pt}
	\begin{equation*}
		\mu = e \langle g|\hat{x}|e\rangle
	\end{equation*} \\
	\hline
	\(P\) & Polarization density:
	\setlength{\abovedisplayskip}{0pt}
	\setlength{\belowdisplayskip}{8pt}
	\begin{equation*}
		P = -e N \langle x \rangle = - N (d \mu^* + d^* \mu)
	\end{equation*} \\
	\hline
	\(P(\omega_0)\) & Positive frequency component of \(P\):
	\setlength{\abovedisplayskip}{0pt}
	\setlength{\belowdisplayskip}{8pt}
	\begin{equation*}
		P(\omega_0) = i \dfrac{N |\mu|^2 w_0 T_2}{\hbar} \hat{L}_2 A \left[ 1 + \dfrac{1}{2 P_s} \hat{L}_1 \left( A^* \hat{L}_2 A + A (\hat{L}_2 A)^* \right) \right]^{-1}
	\end{equation*} \\
	\hline
	
	\multicolumn{2}{|l|}{\textbf{Electric Field Variables and Envelopes}} \\
	\hline
	\(E_f\) & Total electric field:
	\setlength{\abovedisplayskip}{0pt}
	\setlength{\belowdisplayskip}{8pt}
	\begin{equation*}
		E_f(t) = E_a(t) + E_a^*(t)
	\end{equation*} \\
	\hline
	\(E_a\) & Positive frequency component of the electric field:
	\setlength{\abovedisplayskip}{0pt}
	\setlength{\belowdisplayskip}{8pt}
	\begin{equation*}
		E_a(t) = A(t) e^{i \omega_0 t}
	\end{equation*} \\
	\hline
	\(A(t)\) & Slowly varying envelope of the electric field \\
	\hline
	\(E_\pm(z,t)\) & Slowly varying envelopes of forward (\(+\)) and backward (\(-\)) propagating waves \\
	\hline
	\(E(z,t)\) & Extended cavity field unifying forward and backward propagating waves:
	\setlength{\abovedisplayskip}{0pt}
	\setlength{\belowdisplayskip}{8pt}
	\begin{equation*}
		E(z,t) = \begin{cases} E_+(z,t) & \text{for } z \geq 0 \\ E_-(-z,t) & \text{for } z < 0 \end{cases}
	\end{equation*} \\
	\hline
	\(F(z,t)\) & Field normalized with respect to the CW field squared:
	\setlength{\abovedisplayskip}{0pt}
	\setlength{\belowdisplayskip}{8pt}
	\begin{equation*}
		E(z,t) = \sqrt{K(z)} F(z,t)
	\end{equation*} \\
	\hline
	\(f(z,t)\) & Dimensionless field:
	\setlength{\abovedisplayskip}{0pt}
	\setlength{\belowdisplayskip}{8pt}
	\begin{equation*}
		f(z,t) = \dfrac{F(z,t)}{\sqrt{P_0}}
	\end{equation*} \\[0.05mm]
	\hline
	\(f_-(z,t)\) & Counterpropagating field:
	\setlength{\abovedisplayskip}{0pt}
	\setlength{\belowdisplayskip}{8pt}
	\begin{equation*}
		f_-(z,t) = f(-z,t)
	\end{equation*} \\
	\hline
	
	\multicolumn{2}{|l|}{\textbf{Material and Medium Parameters}} \\
	\hline
	\(N\) & Doping concentration (number density of two-level atoms) \\
	\hline
	\(n_0\) & Refractive index of the medium \\
	\hline
	\(\chi^{(3)}\) & Third-order nonlinear susceptibility \\
	\hline
	\(n_2\) & Nonlinear refractive index coefficient (Kerr coefficient):
	\setlength{\abovedisplayskip}{0pt}
	\setlength{\belowdisplayskip}{8pt}
	\begin{equation*}
		n_2 = \dfrac{3 \chi^{(3)} }{4 n_0^2 c \varepsilon_0}
	\end{equation*} \\
	\hline
	\(\gamma_K\) & Kerr nonlinearity coefficient:
	\setlength{\abovedisplayskip}{0pt}
	\setlength{\belowdisplayskip}{8pt}
	\begin{equation*}
		\gamma_K = \dfrac{n_2 \omega_0}{2 n_0 \varepsilon_0}
	\end{equation*} \\
	\hline
	\(P_s\) & Saturation intensity:
	\setlength{\abovedisplayskip}{0pt}
	\setlength{\belowdisplayskip}{8pt}
	\begin{equation*}
		P_s = \left( \dfrac{4 T_1 T_2}{\hbar^2} | \mu |^2 \right)^{-1}
	\end{equation*} \\
	\hline
	\(g_0\) & Small-signal power gain of the medium:
	\setlength{\abovedisplayskip}{0pt}
	\setlength{\belowdisplayskip}{8pt}
	\begin{equation*}
		g_0 = \dfrac{N w_0 |\mu|^2 \omega_0 T_2}{n_0 \varepsilon_0 c \hbar}
	\end{equation*} \\
	\hline
	
	\multicolumn{2}{|l|}{\textbf{Wave Propagation Variables}} \\
	\hline
	\(v_g\) & Group velocity \\
	\hline
	\(\omega_0\) & Carrier angular frequency \\
	\hline
	\(\omega_{eg}\) & Transition angular frequency:
	\setlength{\abovedisplayskip}{0pt}
	\setlength{\belowdisplayskip}{8pt}
	\begin{equation*}
		\omega_{eg} = \dfrac{E_e - E_g}{\hbar}
	\end{equation*} \\
	\hline
	\(k_0\) & Carrier wavevector:
	\setlength{\abovedisplayskip}{0pt}
	\setlength{\belowdisplayskip}{8pt}
	\begin{equation*}
		k_0 = \dfrac{\omega_0 n_0}{c}
	\end{equation*} \\
	\hline
	\(\nlgainvar_\text{NL}\) & Nonlinear gain term at \(\omega_0\):
	\setlength{\abovedisplayskip}{0pt}
	\setlength{\belowdisplayskip}{8pt}
	\begin{equation*}
		\nlgainvar_\text{NL} = \dfrac{g_0}{2} \hat{L}_2 \left[ 1 + \dfrac{1}{P_s} \hat{L}_1 \operatorname{Re} \left( A^* \hat{L}_2 A \right) \right]^{-1} A
	\end{equation*} \\
	\hline
	\(\nlgainvar_\text{NL}^\pm\) & Nonlinear gain action on field\\
	\hline
	
	\multicolumn{2}{|l|}{\textbf{Power and Intensity Variables}} \\
	\hline
	\(P_\pm\) & CW field squared in real space \\
	\hline
	\(P(z)\) & CW field squared in extended space\\
	\hline
	\(P_0\) & CW field squared at \(z = 0\):
	\setlength{\abovedisplayskip}{0pt}
	\setlength{\belowdisplayskip}{8pt}
	\begin{equation*}
		P_0 = P(z \to 0^+)
	\end{equation*} \\
	\hline
	\(K(z)\) & Output coupler-normalized CW power:
	\setlength{\abovedisplayskip}{0pt}
	\setlength{\belowdisplayskip}{8pt}
	\begin{equation*}
		K(z) = \dfrac{P(z)}{P_0}
	\end{equation*} \\
	\hline
	\(K_s\) & Output coupler-normalized saturation power:
	\setlength{\abovedisplayskip}{0pt}
	\setlength{\belowdisplayskip}{8pt}
	\begin{equation*}
		K_s = \dfrac{P_s}{P_0}
	\end{equation*} \\
	\hline
	\(k(z)\) & Saturation-normalized CW power:
	\setlength{\abovedisplayskip}{0pt}
	\setlength{\belowdisplayskip}{8pt}
	\begin{equation*}
		k(z) = \dfrac{K(z)}{K_s}
	\end{equation*} \\
	\hline
	\(k_-(z)\) & Normalized backward wave power:
	\setlength{\abovedisplayskip}{0pt}
	\setlength{\belowdisplayskip}{8pt}
	\begin{equation*}
		k_-(z) = k(-z)
	\end{equation*} \\
	\hline
	\(I_\pm\) & Wave intensity
	\setlength{\abovedisplayskip}{0pt}
	\setlength{\belowdisplayskip}{8pt}
	\begin{equation*}
		I_\pm = 2 n_0 c \varepsilon_0 |E_\pm|^2
	\end{equation*} \\
	\hline
	
	\multicolumn{2}{|l|}{\textbf{Saturation and Gain Terms}} \\
	\hline
	\(D_0\) & Saturation-like term related to population dynamics:
	\setlength{\abovedisplayskip}{0pt}
	\setlength{\belowdisplayskip}{8pt}
	\begin{equation*}
		D_0 = 1 + k s_{++} + k_- s_{--}
	\end{equation*} \\
	\hline
	\(D_2^\pm\) & Saturation-like terms related to coherence dynamics:
	\setlength{\abovedisplayskip}{0pt}
	\setlength{\belowdisplayskip}{8pt}
	\begin{equation*}
		D_2^\pm = 2 \sqrt{k k_-} \, s_{-+}
	\end{equation*} \\
	\hline
	\(S_{ij}\) & Intensity functions associated with saturation:
	\setlength{\abovedisplayskip}{0pt}
	\setlength{\belowdisplayskip}{8pt}
	\begin{equation*}
		S_{ij} = \dfrac{1}{2} \hat{L}_1 \left( F_i^* \hat{L}_2 F_j + F_j \left( \hat{L}_2 F_i \right)^* \right)
	\end{equation*} \\
	\hline
	\(s_{ij}\) & Intensity functions associated with saturation (normalized):
	\setlength{\abovedisplayskip}{0pt}
	\setlength{\belowdisplayskip}{8pt}
	\begin{equation*}
		s_{ij} = \dfrac{1}{2} \hat{L}_1 \left( f_i^* \hat{L}_2 f_j + f_j \left( \hat{L}_2 f_i \right)^* \right)
	\end{equation*} \\
	\hline
	\(s\) & Intensity function:
	\setlength{\abovedisplayskip}{0pt}
	\setlength{\belowdisplayskip}{8pt}
	\begin{equation*}
		s = \operatorname{Re} \hat{\mathcal{L}}_1 \left( f^* \hat{\mathcal{L}}_2 f \right)
	\end{equation*} \\
	\hline
	\(g_{cw}(z)\) & Spatially-dependent gain of CW light
	\setlength{\abovedisplayskip}{0pt}
	\setlength{\belowdisplayskip}{8pt}
	\begin{equation*}
		g_{cw}(z) = g_0 \dfrac{1}{2k} \left( 1 + \dfrac{k - k_- - 1}{\sqrt{(1 + k + k_- )^2 - 4 k k_-}} \right)
	\end{equation*} \\
	\hline
	
	\multicolumn{2}{|l|}{\textbf{Operators and Filters}} \\
	\hline
	\(\hat{L}_i\) & Lorentzian operator associated with \(T_i\):
	\setlength{\abovedisplayskip}{0pt}
	\setlength{\belowdisplayskip}{8pt}
	\begin{equation*}
		\hat{L}_1 = \left( 1 + T_i \dfrac{\partial}{\partial t} \right)^{-1}
	\end{equation*} \\
	\hline
	\(\hat{\mathcal{L}}_i\) & Spatial-domain Lorentzian operator
	\setlength{\abovedisplayskip}{0pt}
	\setlength{\belowdisplayskip}{8pt}
	\begin{equation*}
		\hat{\mathcal{L}}_i = \left( 1 - v_g T_i \dfrac{\partial}{\partial z} \right)^{-1}
	\end{equation*} \\
	\hline
	\(\hat{B}_{in}\) & Higher order Lorentzian (time-domain)
	\setlength{\abovedisplayskip}{0pt}
	\setlength{\belowdisplayskip}{8pt}
	\begin{equation*}
		\hat{B}_{in} =  (\hat{{L}}^{-1}_i-1)^n \hat{{L}}^{n+1}_i
	\end{equation*} \\
	\hline
	\(\hat{\mathcal{B}}_{in}\) & Higher order Lorentzian (spatial-domain)
	\setlength{\abovedisplayskip}{0pt}
	\setlength{\belowdisplayskip}{8pt}
	\begin{equation*}
		\hat{\mathcal{B}}_{in} =  (\hat{\mathcal{L}}^{-1}_i-1)^n \hat{\mathcal{L}}^{n+1}_i
	\end{equation*} \\
	\hline
	
	\multicolumn{2}{|l|}{\textbf{Mean-Field Terms}} \\
	\hline
	\(\tilde{k}[f](z)\) & Convolution with \(k\):
	\setlength{\abovedisplayskip}{0pt}
	\setlength{\belowdisplayskip}{8pt}
	\begin{equation*}
		\tilde{k}[f](z) = \dfrac{1}{L_r} \int_0^{L_r} k(u) f(z + 2u) du
	\end{equation*} \\
	\hline
	\(\langle k \rangle\) & Mean value of \(k\):
	\setlength{\abovedisplayskip}{0pt}
	\setlength{\belowdisplayskip}{8pt}
	\begin{equation*}
		\langle k \rangle = \dfrac{1}{L_r} \int_0^{L_r} k(u) du
	\end{equation*} \\[0.1mm]
	\hline
	
	\multicolumn{2}{|l|}{\textbf{Cavity Parameters}} \\
	\hline
	\(\alpha(z)\) & Spatially-varying loss coefficient \\
	\hline
	\(L_c\) & Physical cavity length \\
	\hline
	\(L_r\) & Cavity round-trip length:
	\setlength{\abovedisplayskip}{0pt}
	\setlength{\belowdisplayskip}{8pt}
	\begin{equation*}
		L_r = 2 L_c
	\end{equation*} \\
	\hline
	\(T_r\) & Cavity round-trip time:
	\setlength{\abovedisplayskip}{0pt}
	\setlength{\belowdisplayskip}{8pt}
	\begin{equation*}
		T_r = \dfrac{L_r}{v_g}
	\end{equation*} \\
	\hline
	
\end{longtable}

\endgroup

\end{document}